\documentclass[acmtog,screen]{acmart}
\AtBeginDocument{%
  }

\setcopyright{cc}
\copyrightyear{2026}
\acmYear{2026}
\acmDOI{10.1145/XXXXXXX.XXXXXXX}
\setcctype[4.0]{by-nc-sa}
\acmJournal{TOG}

\usepackage{booktabs}
\usepackage{multirow}
\usepackage{xcolor}
\usepackage{cuted}
\usepackage{capt-of}

\makeatletter
\usepackage{etoolbox}
\patchcmd{\NAT@citex}
  {\@citea\NAT@hyper@{%
             \NAT@nmfmt{\NAT@nm}%
             \hyper@natlinkbreak{%
               \NAT@aysep\NAT@spacechar}{\@citeb\@extra@b@citeb
             }%
             \NAT@date
           }}
  {\@citea\NAT@nmfmt{\NAT@nm}%
             \NAT@aysep\NAT@spacechar
             \NAT@hyper@{\NAT@date}}
  {}{}
\patchcmd{\NAT@citex}
  {\@citea\NAT@hyper@{%
             \NAT@nmfmt{\NAT@nm}%
             \hyper@natlinkbreak{\NAT@spacechar\NAT@@open\if*#1*\else#1\NAT@spacechar\fi}%
               {\@citeb\@extra@b@citeb}%
             \NAT@date
           }}
  {\@citea\NAT@nmfmt{\NAT@nm}%
             \NAT@spacechar\NAT@@open\if*#1*\else#1\NAT@spacechar\fi
             \NAT@hyper@{\NAT@date}}
  {}{}
\makeatother
\begin{document}
\authorsaddresses{Authors’ Contact Information: Weiyun Jiang, Rice University, Houston, Texas, USA, wyjiang@rice.edu; Haiyun Guo, Rice University, Houston, Texas, USA, hg39@rice.edu; Christopher A. Metzler, University of Maryland College Park, College Park, Maryland, USA, metzler@umd.edu; Ashok Veeraraghavan, Rice University, Houston, Texas, USA, vashok@rice.edu.}
\hypersetup{colorlinks=true, citecolor=ACMDarkBlue, linkcolor=ACMDarkBlue, urlcolor=ACMDarkBlue}
\title{Guidestar-Free Adaptive Optics with Asymmetric Apertures} 
\thanks{Project Webpage: \url{https://weiyunjiang.com/guidestar-free-ao/}}


\author{Weiyun Jiang}
\authornote{Both authors contributed equally to this work.}
\orcid{0000-0002-4078-8133}
\affiliation{%
  \institution{Rice University}
  \city{Houston}
  \state{Texas}
  \country{USA}}
\email{wyjiang@rice.edu}

\author{Haiyun Guo}
\authornotemark[1]
\orcid{0000-0001-6774-0298}
\affiliation{%
  \institution{Rice University}
  \city{Houston}
  \state{Texas}
  \country{USA}}
\email{hg39@rice.edu}

\author{Christopher A.~Metzler}
\authornote{Both authors contributed equally to this work.}
\orcid{0000-0001-6827-7207}
\affiliation{%
  \institution{University of Maryland College Park}
  \city{College Park}
  \state{Maryland}
  \country{USA}}
\email{metzler@umd.edu}
  
\author{Ashok Veeraraghavan}
\authornotemark[2]
\orcid{0000-0001-5043-7460}
\affiliation{%
  \institution{Rice University}
  \city{Houston}
  \state{Texas}
  \country{USA}}
\email{vashok@rice.edu}

\renewcommand{\shortauthors}{Jiang et al.}

\begin{abstract}
This work introduces the first closed-loop adaptive optics (AO) system capable of optically correcting aberrations in real-time without a guidestar or a wavefront sensor. Nearly 40 years ago, Cederquist et al.~demonstrated that asymmetric apertures enable phase retrieval (PR) algorithms to perform fully computational wavefront sensing, albeit at a high computational cost. More recently, Chimitt et al.~extended this approach with machine learning and demonstrated real-time wavefront sensing using only a single (guidestar-based) point-spread-function (PSF) measurement. Inspired by these works, we introduce a {\em guidestar-free} AO framework built around asymmetric apertures and machine learning. Our approach combines three key elements: (1) an asymmetric aperture placed at the system's pupil plane that enables PR-based wavefront sensing, (2) a pair of machine learning algorithms that estimate the PSF from natural scene measurements and reconstruct phase aberrations, and (3) a spatial light modulator that performs optical correction. We experimentally validate this framework on dense natural scenes imaged through unknown obscurants. Our method outperforms state-of-the-art guidestar-free wavefront shaping methods, using an order of magnitude fewer measurements and three orders of magnitude less computation.

\end{abstract}

\ccsdesc[500]{Computing methodologies~Computational photography}
\ccsdesc[500]{Computing methodologies~Reconstruction}

\keywords{Guidestar-free, imaging through obscurants, asymmetric apertures, wavefront shaping, wavefront correction, adaptive optics}
\begin{teaserfigure}
  \includegraphics[width=1.0\linewidth]{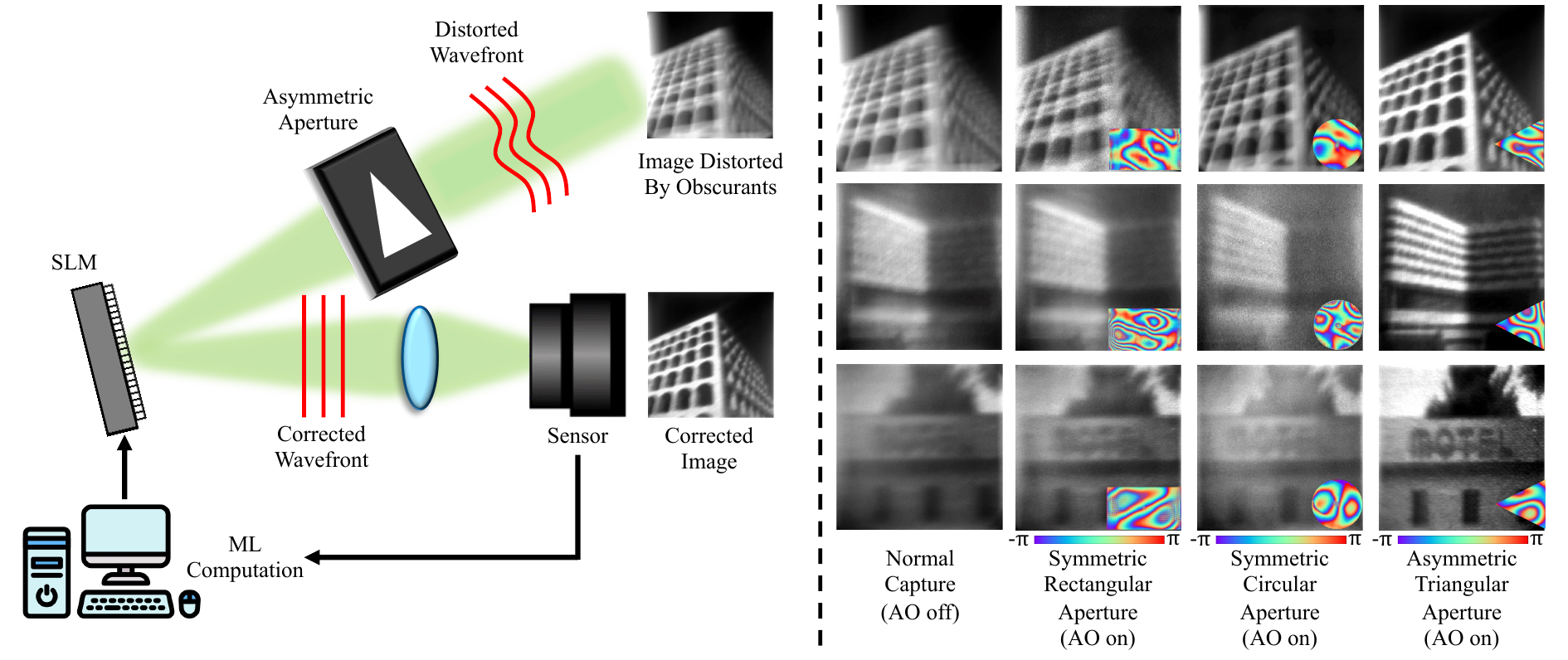}
  \caption{\textbf{Guidestar-free adaptive optics with asymmetric apertures.} 
  Our system combines asymmetric apertures with machine learning to image through severe optical aberrations without wavefront sensors or guidestars. By contrast, conventional symmetric apertures suffer from the conjugate-flip ambiguity, which makes computational wavefront sensing---and by extension AO---ineffective.
  {The above results were experimentally captured with a tabletop prototype.} 
  }
  \Description{The figure has two parts separated by a dashed vertical line. 
    The left part is a system diagram showing the adaptive optics pipeline: 
    a distorted wavefront arrives from a scene obscured by aberrating materials, 
    passes through an asymmetric triangular aperture, and is modulated by a 
    spatial light modulator (SLM) whose correction pattern is computed by a 
    machine learning algorithm running on a computer. The corrected wavefront 
    is then focused by a lens onto an image sensor, producing a sharp output 
    image. The right part is a four-by-three grid of grayscale images comparing 
    four capture configurations across three scenes imaged through different 
    obscurants. The four columns are: normal capture with AO off, AO on with a 
    symmetric rectangular aperture, AO on with a symmetric circular aperture, 
    and AO on with an asymmetric triangular aperture. The AO-off column shows 
    heavily blurred images. The two symmetric aperture columns show very poor 
    correction, with images remaining heavily blurred and largely 
    indistinguishable from the AO-off results, illustrating the conjugate-flip 
    ambiguity problem. Only the asymmetric triangular aperture column shows 
    substantially improved reconstructions with clearly resolved scene details. 
    Each AO-on image includes a small inset showing the estimated phase map, 
    color-coded from negative pi (blue) to positive pi (red).}
  \label{fig:teaser}
\end{teaserfigure}
\maketitle


\section{Introduction}
\label{sec:intro}
\begin{figure}[!ht]
  \centering
   \includegraphics[width=1.0\linewidth]{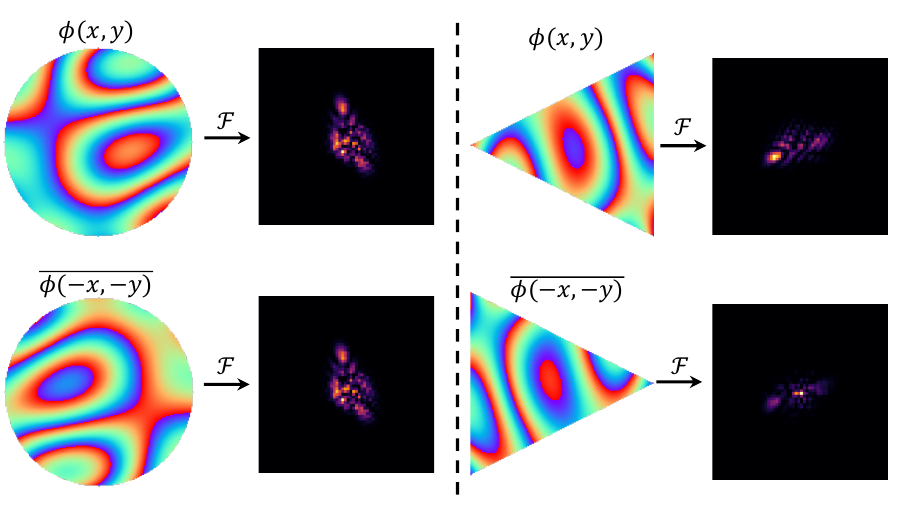}
   \caption{\textbf{Breaking symmetry with asymmetric triangular apertures enables unique phase retrieval from PSFs.} 
   For a circular aperture, the original phase $\phi(x,y)$ and its conjugate-flipped counterpart $\overline{\phi(-x, -y)}$ produce identical PSFs after the Fourier transform, leading to phase ambiguity in intensity-only measurements. In contrast, for a triangular aperture---representative of a broader class of asymmetric apertures such as polygons with an odd number of sides---the wavefront phases $\phi(x,y)$ and $\overline{\phi(-x, -y)}$ yield distinct PSFs. This asymmetry enables unique phase retrieval from intensity measurements, resolving the ambiguity present in symmetric apertures.
   }
   \Description{The figure is divided into two halves by a vertical dashed 
    line. The left half shows a symmetric circular aperture. In the top row, 
    a color-coded phase map labeled phi(x,y) is followed by a Fourier 
    transform arrow and the resulting point spread function (PSF) image on a 
    dark background. The bottom row shows the conjugate-flipped phase map, 
    labeled as the complex conjugate of phi(-x,-y), with its corresponding 
    PSF. The two PSFs on the left are visually identical, demonstrating the 
    conjugate-flip ambiguity. The right half repeats the same layout using an 
    asymmetric triangular aperture. The top row shows phi(x,y) on a 
    triangular domain and its resulting PSF, while the bottom row shows the 
    conjugate-flipped phase and its PSF. In this case, the two PSFs are 
    visibly different in both shape and intensity distribution. This contrast 
    demonstrates that symmetric apertures cannot distinguish between a phase 
    and its conjugate flip from intensity measurements alone, while asymmetric 
    apertures break this ambiguity by producing distinct PSFs for each case.}
   \label{fig:toy}
\end{figure}
Optical imaging systems often suffer from wavefront distortions caused by obscurants, including atmospheric turbulence~\cite{wu2022integrated, cai2025temporally, jiang2023nert, chan2023computational}, biological tissues~\cite{alterman2021imaging, ntziachristos2010going, bar2020rendering, alido2024robust}, fog~\cite{satat2018towards,10.1145/3687930,guo2021beyond}, rain~\cite{garg2004detection}, and imperfect lenses~\cite{10.1145/2516971.2516974, sasian2012introduction,10.1145/3474088,mao2025fovea}. These obscurants scatter and distort light, producing blurry images and severely reducing visibility. 
Adaptive optics (AO) is a powerful tool for imaging through obscurants, which works by actively measuring and correcting wavefront distortions. AO is widely used in fields such as astronomy~\cite{rao2024astronomical, roddier1999adaptive, roggemann2018imaging}, biomedical microscopy~\cite{ji2017adaptive, zhang2023adaptive}, communications~\cite{tyson2002bit, tyson1996adaptive}, and surveillance~\cite{bennet2015adaptive, wood1997estimation}. However, existing AO methods require additional specialized hardware~\cite{cha2010shack, poyneer2003scene, Liang:97, Brunner:21} or rely on multiple coded measurements~\cite{feng2023neuws, xie2024wavemo, yeminy2021guidestar, haim2025image, paxman1992joint, mugnier2006phase, gonsalves1982phase}.

{\it Is it possible to build an AO system without any additional hardware or measurements, using only computation?} An optical system's point-spread-function (PSF) is related to its wavefront error through a Fourier magnitude squared relationship. In principle, researchers could directly estimate a system's wavefront error from its PSF by solving a phase retrieval (PR) problem. Unfortunately, for most optical systems, this problem is fundamentally ill-posed; multiple phase distributions can produce identical intensity patterns due to reflection and conjugation symmetries in the Fourier domain~\cite{fienup1982phase, gonsalves1982phase}. 
\citet{cederquist1989wave} demonstrated that an asymmetric aperture, which has no conjugate symmetries, can remove this ambiguity and make the PR problem well-posed. See Fig.~\ref{fig:toy}.
Still, the high computational costs of iterative phase retrieval algorithms have made computational wavefront sensing systems impractical, until recently.

\citet{chimitt2024phase,chimitt2025wavefront} recently introduced a practical fully computational approach to wavefront sensing. Specifically, their framework leverages an asymmetric aperture to enable a computationally efficient neural network to learn a direct mapping from the PSFs to the underlying phase aberrations. However, while their approach eliminates the need for iterative PR, it still requires guidestar-based calibration, where a known point source is used to obtain the PSF before estimating the phase. This reliance on calibrated illumination limits its applicability in uncontrolled environments, where point-source references are generally unavailable. 

In this paper, we introduce and experimentally validate a novel guidestar-free adaptive optics framework (Fig.~\ref{fig:teaser}).
We build on the work of \citet{chimitt2025wavefront}, reformulating wavefront correction as a PR problem with an asymmetric aperture. However, rather than relying on a guidestar to measure the PSF, our method estimates the PSF from the natural scene measurement using a neural network. This approach eliminates the need for a calibration source and enables practical deployment in real-world imaging scenarios. With the PSF estimate in hand, we use a second neural network to reconstruct the aberration's phase error. The conjugate of the recovered phase error is then added to the pattern displayed on a spatial light modulator (SLM) to perform optical correction. Unlike existing methods that require guidestars~\cite{Tao:12, primmerman1991compensation, horstmeyer2015guidestar}, multiple coded measurements~\cite{feng2023neuws,xie2024wavemo}, or specialized wavefront sensors~\cite{cha2010shack, Liang:97, Brunner:21}, our approach achieves high-quality imaging through unknown obscurants in a closed loop---it recursively removes residual aberrations and can fully correct severe aberrations in 2--4 iterations---with minimal computational overhead.

Our specific contributions include: 
\begin{itemize}
\item We propose a novel guidestar-free adaptive optics framework that leverages asymmetric apertures to perform high-quality imaging through obscurants.
\item We compare our method against existing state-of-the-art guidestar-free wavefront shaping and image deblurring methods; our method consistently outperforms them while using an order of magnitude fewer measurements and three orders of
magnitude less computation.
\item We experimentally validate our method on natural scenes by imaging through various real obscurants, including nail polish, onion skin, and optical diffusers.
\end{itemize}

\section{Related Work}
\label{sec:related}
\paragraph{Imaging Through Obscurants.}
Imaging through obscurants is a persistent challenge in many applications, including astronomy, biomedical microscopy, communications, and surveillance. Conventional AO approaches often rely on specialized wavefront sensors, such as SHWSs~\cite{platt2001history, Liang:97, roddier1999adaptive} and interferometers~\cite{angel1994ground}, to obtain direct measurements of phase distortions. SHWSs utilize arrays of microlenses to focus incident wavefronts onto detector arrays, enabling measurement of local wavefront tilts for reconstruction of the entire wavefront. Interferometers~\cite{malacara2007optical} assess wavefront aberrations by generating interference patterns between a reference beam and a test beam. These measured aberrations are then corrected through optical elements such as deformable mirrors (DMs)~\cite{hampson2021adaptive} or spatial light modulators (SLMs)~\cite{booth2014adaptive}. Despite their effectiveness, these methods require dedicated hardware and precise point-source calibration, which restricts their practicality in uncontrolled environments.

To overcome these limitations, recent developments in AO focus on computational techniques that estimate wavefront distortions directly from camera sensor measurements, thus removing the need for guidestars. Phase-diversity based wavefront shaping~\cite{feng2023neuws, xie2024wavemo} applies known phase modulations to acquire multiple intensity images and jointly optimizes for both phase aberrations and the target image. Image-guided wavefront shaping~\cite{yeminy2021guidestar, haim2025image} begins with a raw, uncorrected measurement and iteratively updates the SLM pattern to enhance image contrast and sharpness. Other iterative approaches achieve guidestar-free correction through confocal feedback~\cite{aizik2024non}, incoherent phase conjugation~\cite{Aizik:22}, or matrix-based covariance decomposition~\cite{sunray2025matrix}. Although these computational methods reduce reliance on specialized sensors and guidestars, the computational burden of iterative optimization limits their processing speed, making real-time correction challenging.

In contrast to these works, our method employs direct, feedforward neural networks to estimate wavefront corrections from conventional camera measurements, eliminating both the need for dedicated wavefront sensors and the computational burden of iterative optimization. This enables significantly faster correction for real-world imaging tasks.
\paragraph{Wavefront Sensing with Asymmetric Apertures.} Traditional wavefront sensing methods suffer from ambiguities due to symmetric apertures, leading to multiple indistinguishable solutions. To address this, researchers have introduced asymmetric apertures to break these ambiguities. Almost 40 years ago, \citet{cederquist1989wave} first demonstrated that asymmetric apertures enable PR via iterative optimization, albeit at a high computational cost. \citet{martinache2013asymmetric} later proposed the Asymmetric Pupil Fourier Wavefront Sensor (APF-WFS), which uses a linear model to estimate phase directly under small aberrations, thereby avoiding iterative retrieval. However, this method requires the assumption that phase aberrations remain within the small-angle approximation, limiting its applicability in real scenarios with larger wavefront distortions. More recently, \citet{chimitt2024phase,chimitt2025wavefront} proposed and demonstrated a deep learning-based method that combines an asymmetric pupil with a neural network for fast wavefront estimation, learning a direct mapping from the PSF to the phase aberration. Unlike APF-WFS, this approach imposes no strict assumptions on the magnitude of phase aberrations, making it more adaptable to real-world conditions. While these methods improve PR, they rely on direct point-source references (guidestars), limiting their applicability in uncontrolled natural environments. In contrast, our method eliminates the need for the point-source calibration, allowing fully passive wavefront sensing in uncontrolled environments.
\section{Imaging Through Obscurants}
\label{sec:Model}
\subsection{Forward Model}
In this paper, we assume we are dealing with a spatially invariant forward model, i.e.,~isoplanatic turbulence or scattering restricted to the memory effect region. Under this model, the measurement process can be described by

\begin{equation}
    Y = X \circledast H + \epsilon,
\end{equation}
where  $Y$ denotes the captured sensor measurement, $\circledast$ denotes two-dimensional convolution, $X$ denotes the unknown target scene, $H$ denotes the system's aberrated PSF, and $\epsilon$ accounts for additive noise.

We further assume we are dealing with monochromatic and incoherent illumination, such that the system's PSF is described by
\begin{equation}
\label{eqn:psf_formula}
    H = \left|\mathcal{F}\{A \circ e^{j \phi_o}\}\right|^2,
\end{equation}
where $\mathcal{F}$  denotes the 2D Fourier transform, $\circ$ denotes element-wise multiplication, $A$ denotes the known aperture shape, and $\phi_o$ denotes the unknown phase aberration/wavefront error~\cite{goodman2005introduction}.

The goal of adaptive optics is to estimate and compensate for the phase aberration $\phi_o$. If one were to know $\phi_o$, one could place the conjugate phase $-\phi_o$ on an SLM or deformable mirror array at the pupil plane and optically correct for the aberration:
\begin{equation}
    Y_{corrected} = X \circledast \left| \mathcal{F} \{A \circ e^{j (\phi_o-\phi_o)} \} \right|^2 + \epsilon= X \circledast H_{corr}  + \epsilon,
\end{equation}
where $H_{corr}=\left|\mathcal{F}\{A\}\right|^2$ is the diffraction-limited PSF of the aperture $A$.
\begin{figure*}[!t]
   \includegraphics[width=1.0\linewidth]{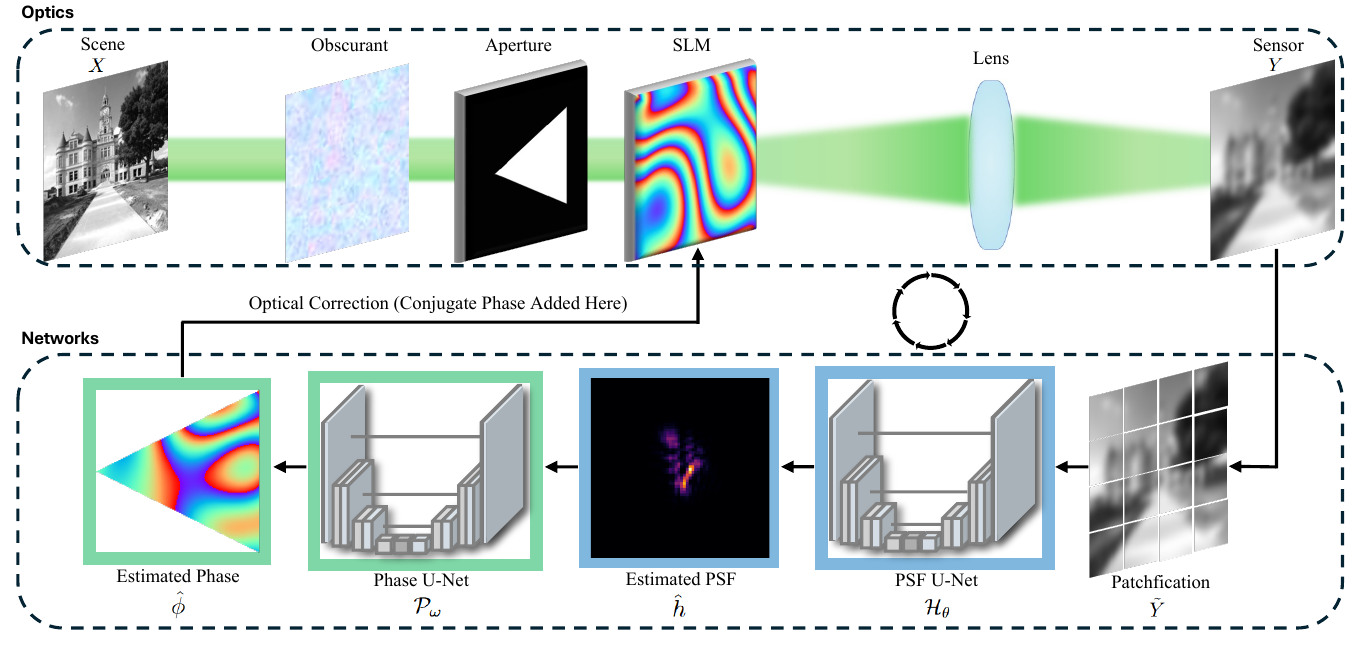}
   \caption{\textbf{Overview of proposed closed-loop guidestar-free AO system.} Light from the scene passes through an obscurant (e.g., nail polish, onion skin, optical diffusers) which introduces an unknown wavefront error $\phi_o$.
   This aberrated wavefront passes through an asymmetric aperture* before reaching a spatial light modulator (SLM) that introduces a phase delay $\phi_{SLM}$.
   A lens forms a blurry and distorted image of the scene; the effective/residual wavefront error of this blurry scene is $\phi=\phi_o+\phi_{SLM}$.
   The PSF U-Net estimates the PSF from the captured image. The Phase U-Net then forms a prediction, $\hat{\phi}$, of the phase error. We add the conjugate of the estimated phase error ($-\hat{\phi}$) to the SLM to optically correct the residual aberration. We repeat the process iteratively until the sensor image is sharp. 
   The results in this manuscript were captured with three AO loops (4 measurements).
   \newline*In practice, the aperture, SLM, and obscurants are co-located at the pupil plane. To realize an asymmetric aperture, we display a checkerboard pattern on a phase-only SLM at the regions where we desire zero amplitude~\cite{mendoza2014encoding}.}
   \Description{The figure is a block diagram divided into two dashed-bordered 
    regions. The top region, labeled Optics, shows the optical path from left to 
    right: a grayscale scene image labeled X, followed by a semi-transparent 
    obscurant, a black asymmetric triangular aperture, a spatial light modulator 
    (SLM) displaying a color-coded phase pattern, a lens, and finally a sensor 
    capturing a blurry grayscale image labeled Y. A green beam illustrates the 
    light path through these components. The bottom region, labeled Networks, 
    shows the computational pipeline flowing from right to left: the captured 
    sensor image is divided into patches (labeled Patchification), which are fed 
    into a PSF U-Net that estimates a point spread function shown as a bright 
    pattern on a dark background. The estimated PSF is then passed to a Phase 
    U-Net, which outputs an estimated phase map displayed as a color-coded image 
    on a triangular aperture domain. An upward arrow connects the estimated 
    phase back to the SLM in the optics region, indicating that the conjugate 
    of the estimated phase is applied to the SLM for optical correction in a 
    closed-loop fashion. The two U-Net architectures are depicted as symmetric 
    encoder-decoder blocks with skip connections.}
   \label{fig:framework}
\end{figure*}
\subsection{Closed-Loop Adaptive Optics}
Now imagine that, either due to errors in the wavefront estimate or changes in the optical aberration over time, the system projects an SLM pattern $\phi_{SLM}\neq-\phi_o$ on the SLM. In this context, an AO system would capture only partially corrected measurements
\begin{equation}
    Y_{partial} = X \circledast \left| \mathcal{F} \{A \circ e^{j \phi} \} \right|^2 + \epsilon,~\text{with}~\phi=\phi_{o}+\phi_{SLM},
\end{equation}
where $\phi$ is the residual, partially-compensated wavefront error in the system. 

A closed-loop adaptive optics system adaptively compensates for this residual wavefront error. That is, the framework computes an estimate, $\hat{\phi}$, and then adds its conjugate to the pattern displayed on the SLM: $\phi_{SLM}^{new}=\phi_{SLM}^{old}-\hat{\phi}$. 

In this work, we develop a closed-loop AO system.

\subsection{Wavefront Sensing as Phase Retrieval}

Before compensating for wavefront errors (residual or not), the system must first estimate them. While this is typically done using a wavefront sensor, it can also be performed by modeling Eq.~\eqref{eqn:psf_formula} as a Fourier phase retrieval problem: Given a measurement of the PSF $H$ and knowledge of the aperture $A$, phase retrieval algorithms can reconstruct the wavefront error $\phi$. 

Note, however, that with symmetric apertures, this problem is fundamentally ill-posed because 
\[
|\mathcal{F}(A_{symm}\circ e^{j\phi})|^2=|\mathcal{F}(A_{symm}\circ {e^{-j\phi_{flip}}})|^2,
\] 
where ${e^{-j\phi_{flip}}}$ is the conjugated and flipped version of $e^{j\phi}$ such that $\phi_{flip}(u, v) = \phi(-u, -v)$. Consequently, the system is non-injective, and two completely different wavefront errors---which call for completely different optical correction---produce identical measurements. 

Fortunately, asymmetric apertures remove this ambiguity and enable PR based wavefront sensing. See~\cite{chimitt2025wavefront} for more information. Fig.~\ref{fig:toy} illustrates the important role asymmetric apertures play in enabling PR-based wavefront sensing. 

\section{Guidestar-Free AO Framework}
\label{sec:methods} 

We combine asymmetric apertures and machine learning to create a powerful closed-loop guidestar-free AO system. 
Our pipeline, illustrated in Fig.~\ref{fig:framework}, images a target scene $X$ through an unknown obscurant. The incoming light is modulated at the pupil plane by an SLM before being imaged onto a camera, which captures a distorted measurement $Y$. (See Sec.~\ref{sec:optical} for a detailed description of the optical setup.)

Given measurements of this form, our AO system operates in three stages. 
In the PSF estimation stage, the (isoplanatically) distorted image $Y$ is broken into patches, which are stacked together as multiple channels and processed by the \textbf{PSF U-Net} $\mathcal{H}_\theta$, which estimates the system's PSF. (Note that this PSF is a function of both the unknown wavefront error $\phi_o$ and the correction pattern $\phi_{SLM}$ currently displayed on the SLM.) 
Next, in the phase error estimation stage, the estimated PSF is passed to the \textbf{Phase U-Net} $\mathcal{P}_\omega$, which predicts the associated (residual) wavefront errors. 
Finally, in the optical correction stage, the conjugate of the wavefront estimate $\hat{\phi}$ is added to the SLM to perform optical correction, and a new image is captured. 

Rather than collapsing the first two stages into a single end-to-end network that maps blurry images directly to wavefront phases, we deliberately decompose the pipeline into PSF estimation followed by phase retrieval. This design is motivated by the fact that the captured image represents spatial intensity at the image plane, while the wavefront phase is defined at the pupil plane---two physically distinct quantities connected through a non-trivial, non-linear relationship governed by Fourier optics. By separating the problem, each sub-network addresses a well-established inverse problem: image-to-PSF estimation operates entirely in the spatial intensity domain, while PSF-to-phase retrieval bridges the image and pupil planes. This decomposition also provides a physically interpretable intermediate output (the PSF) that can be directly inspected, and allows the Phase U-Net to be independently verified using a guidestar, enabling modular diagnosis of each stage.

This three-stage process is repeated until the sensor captures a sharp image or for a fixed number of iterations. 
For simplicity, all experiments in this manuscript were performed with 3 AO loops (4 measurements total).

Details about each stage of our pipeline are provided below.

\subsection{Network Structure}
\textbf{PSF U-Net} $\mathcal{H}_{\theta}$ estimates the PSF from the sensor measurements. Given the original sensor measurement $Y \in \mathbb{R}^{1 \times H \times W}$, we first partition it into patches $\tilde{Y} \in \mathbb{R}^{P \times M \times N}$, where $P$ is the number of patches, and $M$ and $N$ denote the spatial dimensions of each patch. The network then processes these patches and outputs an estimated PSF $\hat{h} \in \mathbb{R}^{1 \times M \times N}$.  

The PSF U-Net $\mathcal{H}_{\theta}$, parameterized by $\theta$, follows a 5-layer encoder-decoder architecture with an initial feature size of 128 and a maximum channel capacity of 4096. Since a standard U-Net preserves input-output spatial dimensions, we reorganize the image into patches and stack them along the channel dimension to effectively reduce the network's spatial output resolution. This allows us to calibrate the patch size to the PSF's specific spatial support, balancing two critical requirements: maintaining a window large enough to capture the PSF's essential tail information for phase recovery, and remaining compact enough to exclude peripheral zero-valued background. Note that the PSF spatial support is known in advance from system calibration, allowing the patch size to be set as a fixed hyperparameter. By concentrating the network's capacity on this relevant region, we prevent the model from wasting parameters on predicting the peripheral low-energy region.

\noindent \textbf{Phase U-Net} $\mathcal{P}_{\omega}$ reconstructs the phase aberration from the estimated PSF. Given an estimated PSF $\hat{h} \in \mathbb{R}^{1 \times M \times N}$, we first apply zero-padding to expand it to $\tilde{h} \in \mathbb{R}^{1 \times H \times W}$, matching the full sensor resolution. The network then processes the padded PSF and predicts the corresponding phase aberration $\hat{\phi} \in \mathbb{R}^{1 \times H \times W}$. For experimental validation, we apply the conjugate phase to the SLM. Since the SLM resolution ($H_{SLM} \times W_{SLM}$) differs from the network output, we resize $\hat{\phi}$ using bilinear interpolation before displaying it on the SLM.

The Phase U-Net $\mathcal{P}_{\omega}$, parameterized by $\omega$, adopts a 5-layer encoder-decoder structure with an initial feature size of 32 and a maximum channel capacity of 512.  Unlike the PSF U-Net, this network requires significantly fewer feature channels since asymmetric apertures break the phase retrieval ambiguity, making phase reconstruction a well-conditioned problem. 

\subsection{Two-Step Network Optimization}
\label{sec:two_step}
We optimize our network in two steps to ensure stable convergence and effective learning of both PSF estimation and phase reconstruction. While the pipeline performs iterative correction during inference, the training procedure for both networks is strictly feedforward, with no feedback or iterative updates between iterations. Each training sample is processed by the networks in a single forward pass, without any iterative refinement. Iterative correction is applied only at inference time, not during training.



\noindent \textbf{Step 1: Independent Training.}  
In the first independent training step, we train the PSF U-Net $\mathcal{H}_{\theta}$ and the Phase U-Net $\mathcal{P}_{\omega}$ separately. The PSF U-Net is optimized using an $L_2$ loss to minimize the reconstruction error between the predicted and ground truth PSFs, following:
\begin{equation}
    \min_{\theta} \| \mathcal{H}_{\theta}(\tilde{Y}) - h_{gt} \|_2^2, 
\end{equation}
where $h_{gt}$ is the ground truth PSF.

Meanwhile, the Phase U-Net $\mathcal{P}_{\omega}$ is trained using a gradient-based loss to enforce smoothness and preserve local variations in the reconstructed phase. Since absolute phase values may shift arbitrarily (e.g., adding a constant to all phase values does not affect the physical wavefront), we minimize the discrepancy between the phase gradients rather than absolute values:
\begin{equation}
\label{eqn:grad}
\min_{\omega} \left\| \nabla_x \mathcal{P}_{\omega}(\tilde{h}) - \nabla_x \phi_{gt} \right\|^2 + \left\| \nabla_y \mathcal{P}_{\omega}(\tilde{h}) - \nabla_y \phi_{gt}  \right\|^2,
\end{equation}
where $\nabla_x$ and $\nabla_y$ represent horizontal and vertical gradient respectively, and $\phi_{gt}$ denotes ground truth phase.

\noindent \textbf{Step 2: Joint Optimization.}  
In the second joint optimization step, we fine-tune the entire network by training the Phase U-Net $\mathcal{P}_{\omega}$ while keeping the PSF U-Net $\mathcal{H}_{\theta}$ fixed. By freezing $\mathcal{H}_{\theta}$, we ensure that the learned PSF estimation remains stable while allowing $\mathcal{P}_{\omega}$ to refine phase predictions based on more stable PSF inputs. This step improves the consistency between PSF estimation and phase reconstruction. We continue optimizing $\mathcal{P}_{\omega}$ using the same gradient-based loss (Eq.~(\ref{eqn:grad})).

\subsection{Training \& Datasets}  
Networks were trained using entirely simulated data. All training and experiments in this work are implemented in PyTorch and conducted on a single NVIDIA A100 GPU. Although our proposed pipeline operates iteratively during inference (see Fig.~\ref{fig:framework}), the network is trained in a feedforward manner using two-step optimization, as described in Sec.~\ref{sec:two_step}.

The training procedure consists of two steps. In the first step, the Phase U-Net and PSF U-Net are optimized separately using the Adam optimizer with $\beta=(0.9, 0.999)$. To train the PSF U-Net, we use 100{,}000 images from the MIT Places2 dataset~\cite{zhou2017places} and a total of 7.8 million PSFs. We set the learning rate to $2 \times 10^{-4}$ and train the network for approximately five days. During training, we randomly sample an image and a PSF to generate blurred image–PSF pairs as input. Each image is resized to $256 \times 256$ and divided into sixteen $64 \times 64$ patches. To train the Phase U-Net, we generate a paired phase–PSF dataset by simulating random phase aberrations using the first 36 modes of Zernike polynomials, where the coefficients are sampled from independent zero-mean normal distributions with standard deviations drawn between 0.01 and 1.5. The corresponding PSFs are computed by taking the squared magnitude of the Fourier transform of the complex pupil function, where the pupil function is defined by the chosen aperture and the simulated phase aberrations, following Eq.~(\ref{eqn:psf_formula}). The Phase U-Net is trained for approximately one day with a learning rate of $2 \times 10^{-5}$.

In the second step, we freeze the weights of the PSF U-Net and fine-tune the Phase U-Net for an additional two days. The Phase U-Net takes as input the estimated PSF produced by the frozen PSF U-Net.
\begin{figure*}[!ht]
    \includegraphics[width=1.0\linewidth]{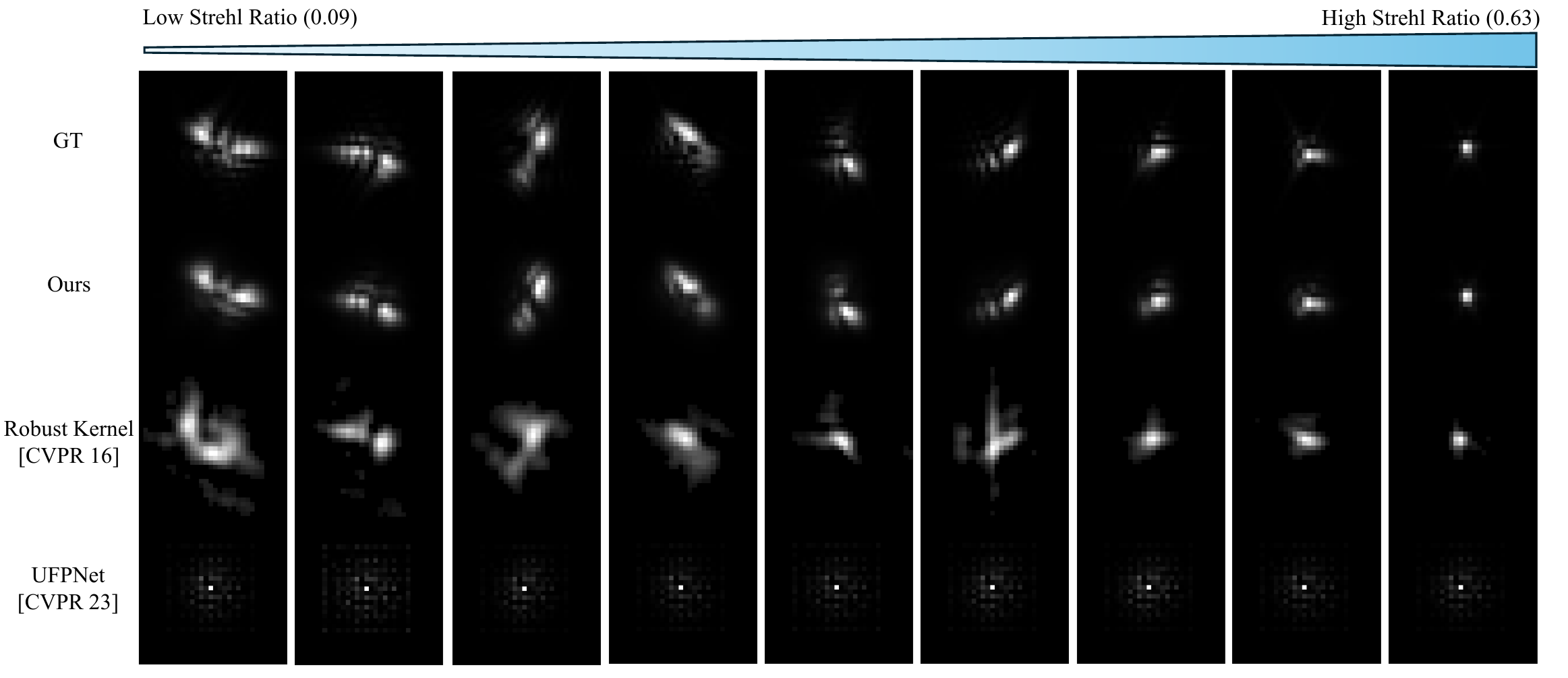}
    \caption{\textbf{Estimated and ground-truth PSFs.}
    Each column presents a PSF example, arranged in order of increasing Strehl ratio from left to right. The top row shows the ground-truth PSFs, followed by estimates from our method, Robust Kernel~\cite{pan2016robust}, and UFPNet~\cite{fang2023self}. We use the official open-source implementation of UFPNet and the released code of Robust Kernel. Robust Kernel and UFPNet provide sub-optimal kernel estimates because they are designed primarily for image deblurring.} 
    \Description{The figure displays a grid of point spread function (PSF) 
    images organized into four rows and nine columns. A gradient color bar at 
    the top spans from light blue on the left labeled Low Strehl Ratio (0.09) 
    to dark blue on the right labeled High Strehl Ratio (0.63), indicating 
    that columns are arranged by increasing Strehl ratio from left to right. 
    The four rows are labeled from top to bottom: GT (ground truth), Ours, 
    Robust Kernel [CVPR 16], and UFPNet [CVPR 23]. All PSF images are shown 
    as bright patterns on dark backgrounds. The ground truth PSFs transition 
    from large, complex, spread-out patterns at low Strehl ratios to smaller, 
    more compact patterns at high Strehl ratios. The proposed method's 
    estimates closely match the ground truth across the range. Robust Kernel 
    produces roughly plausible shapes but with noticeable structural 
    inaccuracies. UFPNet produces near-delta-function outputs that fail to 
    capture the PSF structure, appearing as small bright dots with faint 
    grid-like artifacts across all columns regardless of Strehl ratio.}
\label{fig:psf_est}
\end{figure*}
\subsection{Inference} During the experiments, we perform 3 adaptive optics loops (4 measurements total). 
Most optical aberrations are corrected in the first two loops, with subsequent loops further reducing residual errors.

\section{Results}
\label{sec:results}
We evaluate the proposed method in two main aspects: the accuracy of PSF estimation in simulation, and the effectiveness of wavefront shaping in both simulated environments and real-world imaging through obscurants.

\subsection{Simulated Results}
\subsubsection{Accuracy of PSF Estimation}
To evaluate the accuracy of our estimated PSFs, we construct a hold-out test set consisting of 3,000 images from the MIT Places2~\cite{zhou2017places} dataset and 3,000 previously unseen PSFs generated using Zernike polynomials. The test blurred images are synthesized by convolving the clean test images with these PSFs. We use these blurred images to evaluate the accuracy of PSF estimation of our proposed PSF U-Net. For comparison, we include the following baseline methods:
\begin{itemize}
    \item \textbf{Robust Kernel}~\cite{pan2016robust}: A non-deep-learning blind deblurring method that removes outliers (such as saturation and non-Gaussian noise) from salient edge selection to enable accurate kernel estimation and restoration.
    \item \textbf{UFPNet}~\cite{fang2023self}: A deep learning-based image deblurring method that incorporates non-uniform kernel estimation via normalizing flow-based motion priors and uncertainty learning.
\end{itemize}
Note that to date, kernel estimation methods have been primarily designed to enable image deblurring; therefore, their kernel estimation performance may be suboptimal when directly evaluated for PSF recovery.
\begin{table}[!t]
\caption{\textbf{Quantitative comparison of PSF estimation accuracy.} The proposed method quantitatively outperforms Robust Kernel~\cite{pan2016robust} and UFPNet~\cite{fang2023self} in terms of PSNR and SSIM. Robust Kernel and UFPNet provide sub-optimal kernel estimates because they are designed primarily for image deblurring. PSNR and SSIM are computed over the central cropped region of each PSF to focus evaluation on the energy-concentrated region and avoid background domination. \textbf{Bold} indicates the best result.}
\label{tab:quan_psf}
\begin{tabular}{lcc}
\toprule
Method & PSNR $\uparrow$ & SSIM $\uparrow$ \\
\midrule
UFPNet {[}CVPR 2023{]}        & 17.23 & 0.6359 \\
Robust Kernel {[}CVPR 2016{]} & 24.42 & 0.8960 \\
Ours                          & \textbf{39.22} & \textbf{0.9891} \\
\bottomrule
\end{tabular}
\end{table}
In Fig.~\ref{fig:psf_est}, our method produces PSF estimates that are visually closest to the ground truth. Table~\ref{tab:quan_psf} reports the PSNR and SSIM of the estimated PSFs for our method, Robust Kernel, and UFPNet, providing a quantitative comparison of estimation accuracy. To avoid artificially inflating PSNR due to the large number of near-zero pixels in the 
background, all metrics are computed over the central cropped $64 \times 64$ region of the PSF, where most of the energy is concentrated.
\begin{figure*}[!ht]
    \includegraphics[width=1.0\linewidth]{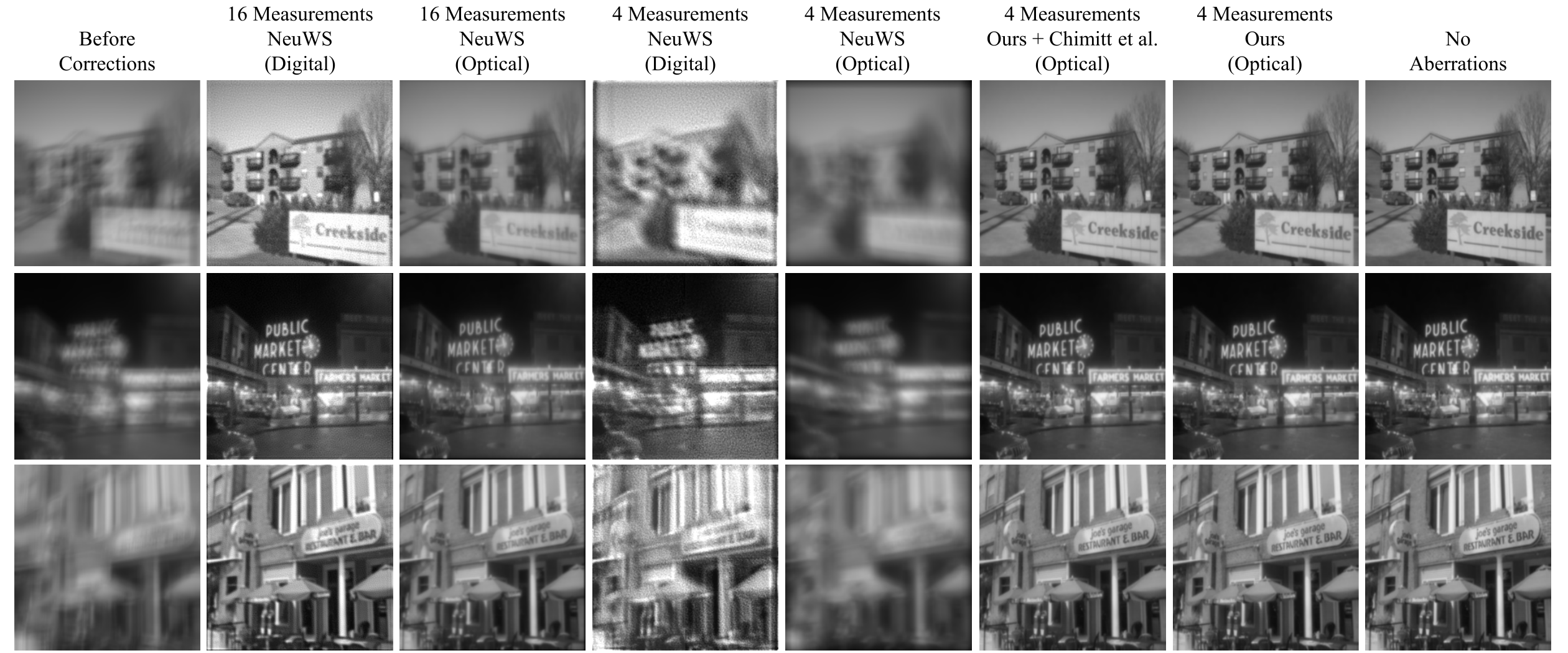}
    \caption{
    \textbf{Simulated results for guidestar-free wavefront shaping compared with NeuWS.}
    We present a qualitative comparison between our method and the state-of-the-art guidestar-free wavefront shaping method, NeuWS~\cite{feng2023neuws}. Each column represents a different method/measurement setting. NeuWS includes both digital and optical correction results, while our method and our method combined with \citet{chimitt2025wavefront} provide only optical correction results. Notably, our approach achieves high-quality image restoration with only four measurements, whereas the performance of NeuWS declines significantly as the number of measurements decreases from 16 to 4. These results demonstrate that our method enables high-quality reconstruction using only four measurements, substantially reducing the measurement requirements compared to NeuWS.
    }
    \Description{The figure presents a grid of grayscale images with three 
    rows and eight columns. Each row shows a different scene: a building with 
    a Creekside sign, a nighttime view of Public Market Center, and a street 
    with storefronts and umbrellas. The eight columns from left to right are: 
    Before Corrections, 16 Measurements NeuWS (Digital), 16 Measurements 
    NeuWS (Optical), 4 Measurements NeuWS (Digital), 4 Measurements NeuWS 
    (Optical), 4 Measurements Ours + Chimitt et al. (Optical), 4 Measurements 
    Ours (Optical), and No Aberrations. The Before Corrections column shows 
    heavily blurred images. NeuWS with 16 measurements produces reasonable 
    results, but its quality degrades substantially when reduced to 4 
    measurements, with images remaining blurry and lacking detail. The 
    proposed method and the proposed method combined with Chimitt et al., both 
    using only 4 measurements, produce sharp results that are visually close 
    to the No Aberrations reference in the rightmost column.}
    \label{fig:qual_ws}
\end{figure*}
\begin{figure*}[!ht]
    \includegraphics[width=0.62\linewidth]{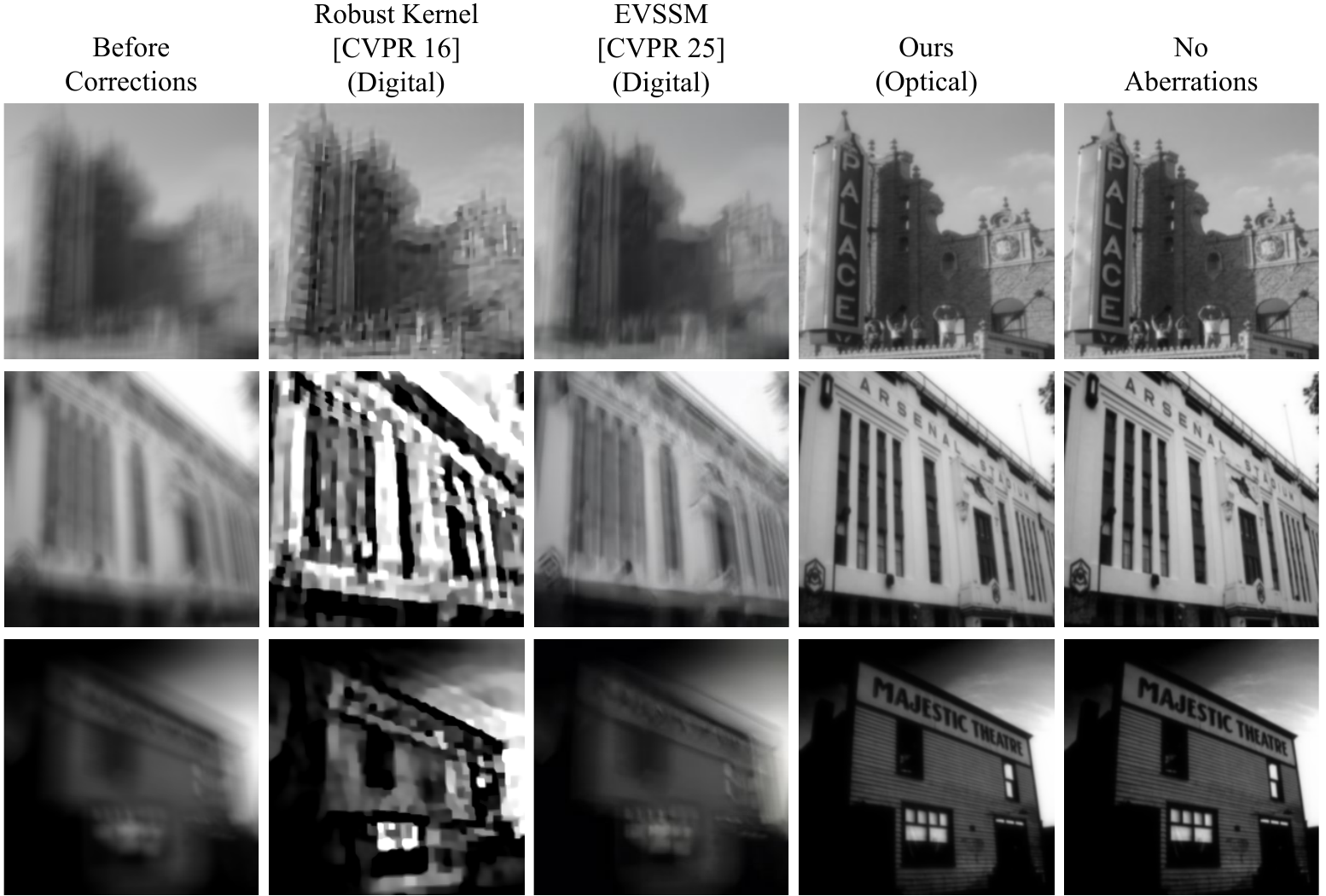}
    \caption{
    \textbf{Simulated comparisons against deblurring baselines.}
    We compare our optical correction approach against two digital deblurring 
    methods, Robust Kernel~\cite{pan2016robust}, a non-deep-learning method, 
    and EVSSM~\cite{kong2025efficient}, a state-of-the-art deep-learning-based 
    method. Each row presents a different scene, and each column displays 
    results from a specific method, along with the uncorrected input and the 
    aberration-free reference. Both digital methods fail to recover meaningful 
    scene detail, producing severe artifacts or residual blur. In contrast, 
    our optical correction successfully restores sharp, high-quality images 
    that closely match the aberration-free reference.
    }
    \Description{
    The figure shows three rows of grayscale images, each row depicting a different scene. Columns correspond to different correction methods: before correction, Robust Kernel (digital correction), EVSSM (digital correction), our method (optical correction), and a reference with no aberrations. The images illustrate that digital correction methods often leave residual artifacts or fail to recover detail, while our optical correction approach produces images that closely resemble the no-aberration reference.
    }
    \label{fig:qual_deconv}
\end{figure*}
\begin{table*}[!ht]
\caption{\textbf{Quantitative evaluation of existing state-of-the-art image deblurring and guidestar-free wavefront shaping methods on both simulated and real experiments.} We report PSNR/SSIM for image quality and computation cost (time and FLOPs). Our method achieves significantly higher image quality with minimal computation, outperforming both digital and optical baselines. \textbf{Bold} indicates the best result, and \underline{underline} indicates the second-best result for each column.}
\label{tab:quan_results}
\begin{tabular}{lccccccc}
\toprule
\multirow{2}{*}{Method} & \multirow{2}{*}{Correction Type} & \multicolumn{2}{c}{Simulation} & \multicolumn{2}{c}{Real} & \multicolumn{2}{c}{Computation Cost} \\ \cmidrule(lr){3-4} \cmidrule(lr){5-6} \cmidrule(lr){7-8}
 & & PSNR $\uparrow$ & SSIM $\uparrow$ & PSNR $\uparrow$ & SSIM $\uparrow$ & Time (sec) $\downarrow$ & FLOPs (G) $\downarrow$ \\ 
\midrule
Robust Kernel & Digital & 15.61 & 0.4062 & 14.75 & 0.2558 & $\sim$90 & \textbf{$\sim$2} \\
\hline
EVSSM & Digital & 18.53 & 0.5027 & {\underline{15.38}} & {\underline{0.2723}} & {\underline{$\sim$0.1}} & $\sim$150 \\
\hline
\multirow{2}{*}{\begin{tabular}[c]{@{}l@{}}NeuWS \\ 4 meas.\end{tabular}} & Digital & 15.33 & 0.3965 & - & - & \multirow{2}{*}{$\sim$15} & \multirow{2}{*}{$\sim$4.4e3} \\
 & Optical & 14.69 & 0.3912 & - & - & & \\
\hline
\multirow{2}{*}{\begin{tabular}[c]{@{}l@{}}NeuWS \\ 16 meas.\end{tabular}} & Digital & 20.97 & 0.6208 & - & - & \multirow{2}{*}{$\sim$20} & \multirow{2}{*}{$\sim$1.66e4} \\
 & Optical & 23.47 & 0.7277 & - & - & & \\
\hline
\multirow{2}{*}{\begin{tabular}[c]{@{}l@{}}WaveMo + NeuWS \\ 16 meas.\end{tabular}} & Digital & 21.13 & 0.6585 & - & - & \multirow{2}{*}{$\sim$20} & \multirow{2}{*}{$\sim$1.7e4} \\
 & Optical & 27.47 & 0.8516 & - & - & & \\
\hline
\multirow{2}{*}{\begin{tabular}[c]{@{}l@{}}NeuWS \\ 100 meas.\end{tabular}} & Digital & 25.58 & 0.8219 & - & - & \multirow{2}{*}{$\sim$85} & \multirow{2}{*}{$\sim$1e5} \\
 & Optical & \textbf{39.23} & \textbf{0.9877} & - & - & & \\
\hline
Ours + Chimitt et al. & Optical & 35.34 & 0.9671 & - & - & \textbf{$\sim$0.05} & {\underline{$\sim$16}} \\
\hline
Ours & Optical & {\underline{37.95}} & {\underline{0.9865}} & \textbf{25.33} & \textbf{0.7421} & \textbf{$\sim$0.05} & {\underline{$\sim$16}} \\
\bottomrule
\end{tabular}
\end{table*}
\subsubsection{Wavefront Shaping Reconstructions}
To evaluate the wavefront shaping performance of our proposed method, we use a separate hold-out test set comprising 3,000 images from the MIT Places2 dataset and 3,000 unseen PSFs generated using Zernike polynomials. The test blurred images are synthesized by convolving the clean test images with these PSFs. Our evaluation follows a three-stage pipeline: (1) estimate the PSF from the blurred images using the proposed PSF U-Net, (2) estimate the associated wavefront error via the Phase U-Net, and (3) perform optical correction by adding the conjugate of the estimated wavefront error. We compare our approach with both state-of-the-art wavefront shaping methods and image deblurring methods.

The wavefront shaping baselines are as follows:
\begin{itemize}
\item \textbf{NeuWS~\cite{feng2023neuws}}: A guidestar-free wavefront shaping method that jointly estimates the wavefront error and the clean image by applying random phase modulations and acquiring multiple modulated measurements. 

NeuWS employs a spatial light modulator (SLM) in the optical path to apply random phase masks across different measurements. Each mask produces a distinct modulated measurement of the same scene. The algorithm alternates between recovering the underlying object and estimating the aberrated wavefront from this set of measurements. In our implementation, we follow the procedure described in the original paper and generate synthetic measurements using Zernike polynomials. For each scene, we sample random phase modulations, apply them to the wavefront, and propagate the result to the sensor plane to form modulated intensity images. These simulated sensor images, together with the applied phase modulations, are then provided as inputs to the iterative reconstruction algorithm of NeuWS.

\item \textbf{WaveMo~\cite{xie2024wavemo}}: A guidestar-free method that extends NeuWS by replacing random phase modulation with learned phase modulation strategies. 

Instead of selecting the modulations independently at random, WaveMo employs a trained modulation generator network that produces phase masks optimized for the intensity measurements. This improves measurement efficiency and substantially reduces the number of required measurements. In our implementation, we apply the learned modulation patterns released with the original work to the input wavefront and propagate them to the sensor plane to generate the measurements. These measurements, together with the corresponding learned modulation patterns, are then provided as inputs to the iterative reconstruction algorithm of NeuWS.

\item \textbf{\citet{chimitt2025wavefront}}: A guidestar-based method that estimates the Zernike coefficients of the wavefront error. 

Chimitt et al. propose a guidestar-based wavefront correction framework that assumes access to a bright point-source guidestar propagating through the same obscurant as the target object. The recorded PSF of the guidestar is used to estimate Zernike coefficients of the wavefront error, which then drive a correction device (e.g., deformable mirror or SLM) to restore image quality. This approach is highly effective when a guidestar is available, but is not applicable in general guidestar-free scenarios.  

To enable comparison under guidestar-free conditions, we adapt Chimitt et al.’s method by replacing the explicit guidestar PSF measurement with the PSF predicted by our PSF U-Net from a blurred intensity measurement. In this hybrid variant, the Zernike coefficient regression of Chimitt et al. remains unchanged; however, its input PSF comes from our network’s estimation rather than the ground-truth guidestar PSF assumed in their work. This adaptation allows us to compare our framework against a strong guidestar-based baseline while ensuring consistent evaluation settings (i.e., no actual guidestar present).
\end{itemize}

In guidestar-free wavefront shaping, the number of measurements corresponds to the number of distinct phase-modulated sensor images captured. Each measurement is obtained by applying a different modulation pattern and recording the resulting intensity image at the detector. More measurements provide richer diversity for wavefront estimation, but at the cost of increased computation and acquisition time.
\begin{figure*}[!ht]
    \includegraphics[width=1.0\linewidth]{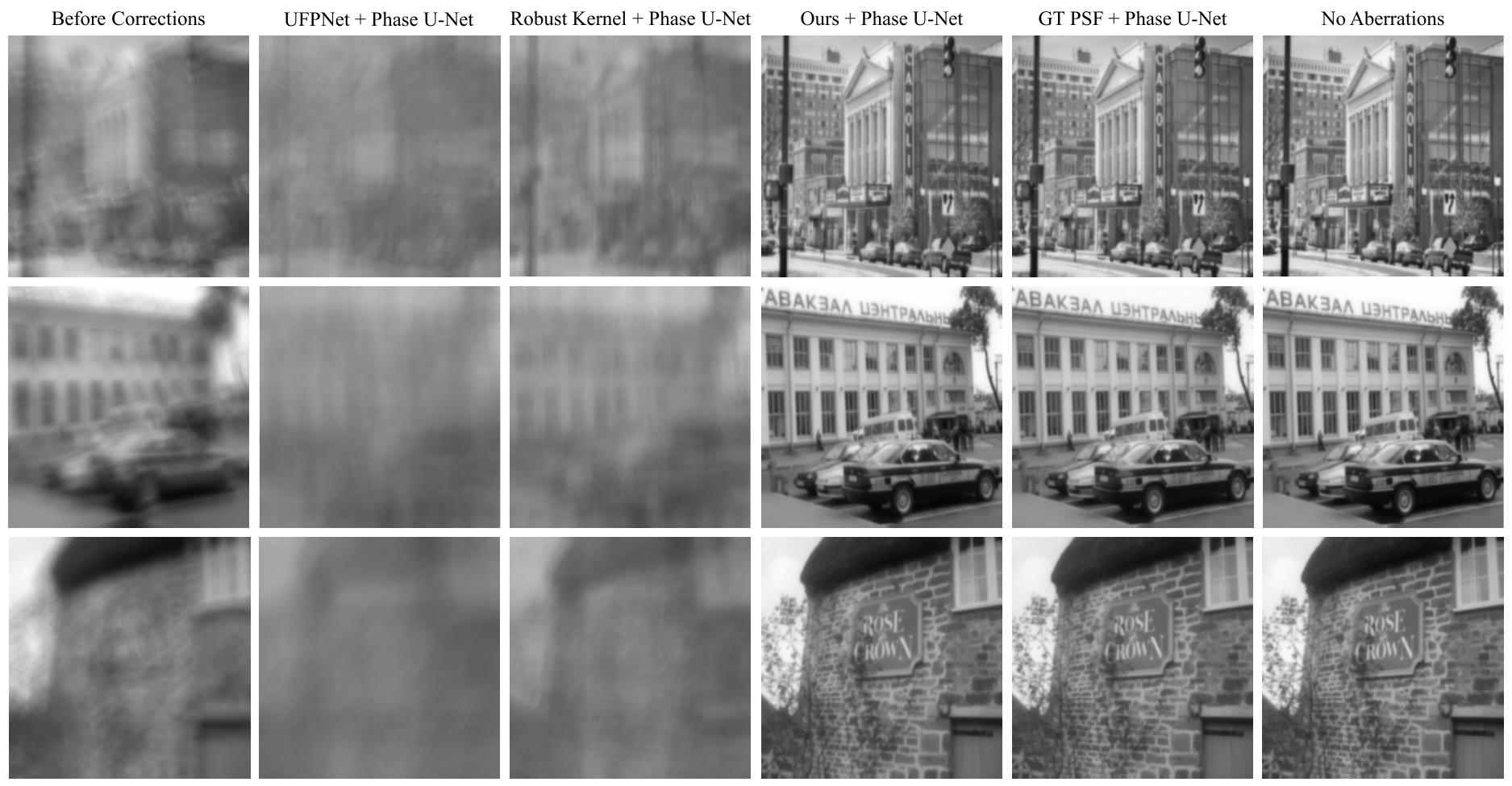}
    \caption{\textbf{Ablation on PSF estimation.} We evaluate the 
    effect of different PSF estimation methods on wavefront correction 
    performance. We compare reconstructions using PSFs estimated by 
    UFPNet~\cite{fang2023self}, Robust Kernel~\cite{pan2016robust}, and our 
    method, along with the ground-truth PSF as an upper bound. Our estimated 
    PSF enables effective correction and produces results comparable to those 
    achieved with the ground-truth PSF, while PSFs estimated by UFPNet or 
    Robust Kernel do not yield satisfactory restoration.}
    \Description{The figure contains a grid of grayscale images arranged in 
    three rows and six columns. Each row corresponds to a different scene: a 
    building with a Carolina sign, a street with a car and a building with 
    Cyrillic text, and a stone wall with a Rose and Crown sign. The six 
    columns from left to right are: Before Corrections, UFPNet + Phase U-Net, 
    Robust Kernel + Phase U-Net, Ours + Phase U-Net, GT PSF + Phase U-Net, 
    and No Aberrations. The Before Corrections column shows blurred images. 
    The UFPNet and Robust Kernel columns remain heavily blurred with no 
    meaningful improvement over the uncorrected input. The proposed method 
    produces sharp, detailed images that are visually close to the GT PSF 
    column, which itself closely matches the No Aberrations reference. This 
    demonstrates that the proposed PSF estimation enables effective wavefront 
    correction comparable to using the ground-truth PSF.}
    \label{fig:psf_ablation}
\end{figure*}
\begin{table}[!ht]
\caption{\textbf{Wavefront shaping performance with different PSF estimation methods.} Our method achieves performance comparable to using the ground-truth PSF. \textbf{Bold} indicates the best result.}
\label{tab:ablation_psf}
\begin{tabular}{l c c}
\toprule
Method & PSNR $\uparrow$ & SSIM $\uparrow$ \\
\midrule
UFPNet + Phase U-Net        & 15.65          & 0.3549 \\
Robust Kernel + Phase U-Net & 16.85          & 0.3954 \\
Ours + Phase U-Net          & \textbf{38.89} & \textbf{0.9884} \\
\midrule
GT PSF + Phase U-Net            & 40.90          & 0.9961 \\
\bottomrule
\end{tabular}
\end{table}
As shown in Table~\ref{tab:quan_results}, NeuWS requires nearly 100 measurements to achieve satisfactory reconstruction quality, and also incurs a high computational cost. With fewer measurements, its performance degrades substantially---at around 16 measurements it produces only reasonable reconstructions, and with 4 measurements the results degrade significantly. These trends are also evident qualitatively in Fig.~\ref{fig:qual_ws}, where NeuWS reconstructions worsen visibly as the number of measurements decreases. In contrast, our method achieves high-quality reconstruction with only four measurements, while requiring orders of magnitude fewer computations and significantly shorter runtime.

The compared image deblurring methods include:
\begin{itemize}
\item \textbf{EVSSM~\cite{kong2025efficient}}: An efficient state space model designed for image deblurring.
\item \textbf{Robust Kernel~\cite{pan2016robust}}: A classical deblurring method that performs iterative kernel estimation and multi-scale deconvolution.
\end{itemize}
As shown in Fig.~\ref{fig:qual_deconv}, both Robust Kernel and EVSSM fail to restore images degraded by strong aberrations. Robust Kernel, a classical iterative deblurring method, leaves heavy artifacts and recovers little fine structure, while EVSSM, despite being a state-of-the-art deep learning model for image deblurring, also struggles and produces results with severe residual blur. In contrast, our proposed method, which performs optical correction, restores images that closely resemble the no-aberration reference. These results highlight that conventional image deblurring approaches, whether classical or learning-based, are not effective for imaging through complex aberrations, whereas our wavefront correction framework successfully recovers high-quality images.
\begin{figure*}[!ht]
    \includegraphics[width=1.0\linewidth]{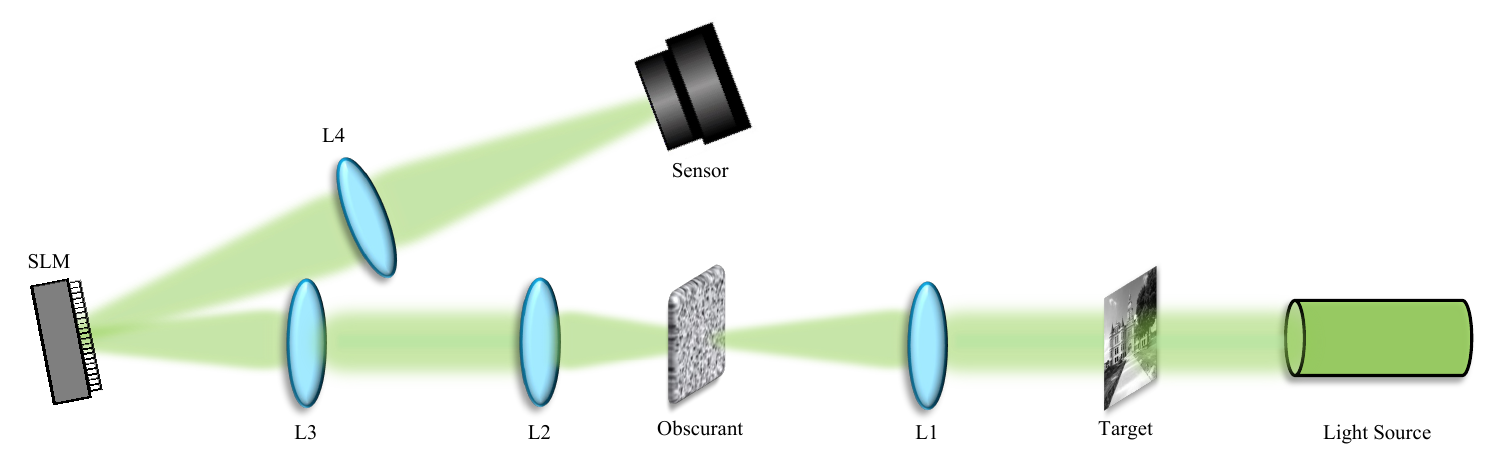}
    \caption{\textbf{Optical setup.} A spatially incoherent light source illuminates the target and passes through the obscurant. The SLM is positioned at the Fourier plane to project the estimated phase patterns. The corrected wavefront is then imaged onto a camera.}
    \Description{The diagram shows a horizontal optical setup with components 
    arranged from right to left. On the far right, a green cylindrical light 
    source illuminates a target displaying a grayscale image of a building. 
    Light passes through lens L1, then through an obscurant, followed 
    by lens L2 and lens L3. On the far left, a spatial light modulator (SLM) 
    is positioned at the Fourier plane. The beam reflects off the SLM and is 
    directed upward through lens L4 onto a sensor at the top of the diagram. 
    A green beam illustrates the light path through all components. All 
    components are labeled: Light Source, Target, L1, Obscurant, L2, L3, SLM, 
    L4, and Sensor.}
    \label{fig:optical_setup}
\end{figure*}

\begin{figure*}[!ht]
    \includegraphics[width=0.70\linewidth]{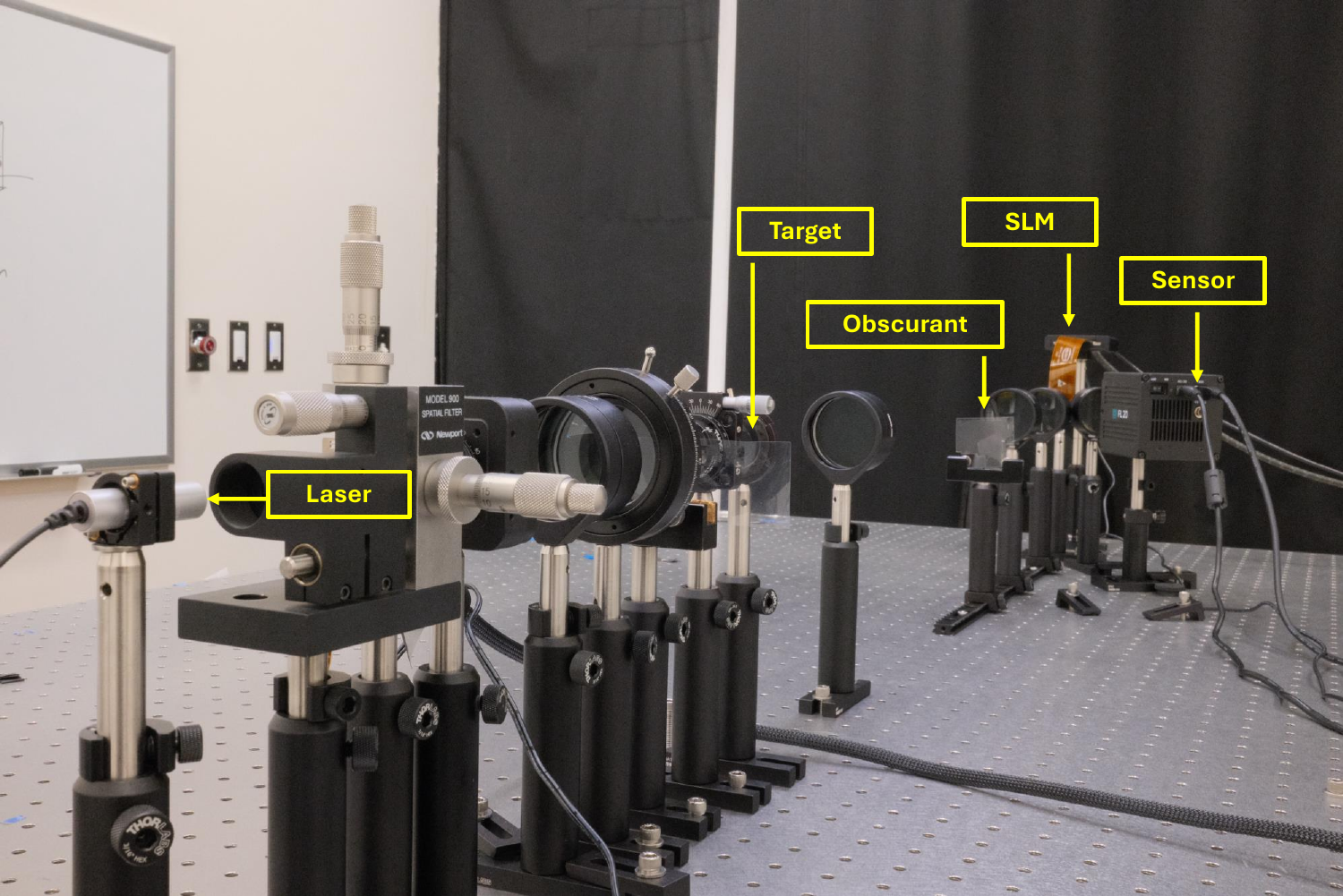}
    \caption{\textbf{Photo of optical setup.} The setup begins with a collimated 520 nm laser that passes through a polarizer and a rotating diffuser. The beam then illuminates the target, a natural scene. The 
    obscurant is positioned at the Fourier plane, and a 4f system relays 
    this plane onto the SLM. The corrected wavefront is then imaged onto 
    the camera sensor.}
    \Description{A photograph of the tabletop optical setup on a metal 
    optical breadboard with a grid of tapped mounting holes. Components are 
    arranged in a line from left to right and labeled with yellow boxes and 
    arrows. On the far left, a laser source is mounted on an adjustable 
    post. Next, a spatial filter assembly and lens tube are visible, followed 
    by the target holder. In the middle, a small transparent obscurant is 
    mounted in a circular holder. Further to the right, an orange-bodied 
    spatial light modulator (SLM) sits on a post mount, and at the far right 
    a camera sensor is mounted. Various lenses on post mounts are positioned 
    between the components. Black curtains hang in the background to block 
    stray light, and cables connect the SLM and sensor for control and data 
    acquisition.}
    \label{fig:photo_optics}
\end{figure*}
Table~\ref{tab:quan_results} further reports the computation cost of each method in terms of time and FLOPs. NeuWS with 100 measurements requires ${\sim}85$ seconds and ${\sim}10^{5}$\,G FLOPs per scene, and even its 4-measurement variant costs ${\sim}15$ seconds and ${\sim}4.4{\times}10^{3}$\,G FLOPs while delivering poor image quality. EVSSM is fast at ${\sim}0.1$ seconds but demands ${\sim}150$\,G FLOPs and, as shown in the same table, produces substantially lower PSNR/SSIM than our approach. Robust Kernel, although low in FLOPs (${\sim}2$\,G), requires ${\sim}90$ seconds due to its iterative optimization. Our method runs in ${\sim}0.05$ seconds with only ${\sim}16$\,G FLOPs---three orders of magnitude faster and fewer FLOPs than NeuWS (100 measurements), making it well suited for real-time wavefront correction.
\begin{figure*}[!ht]
    \includegraphics[width=1.0\linewidth]{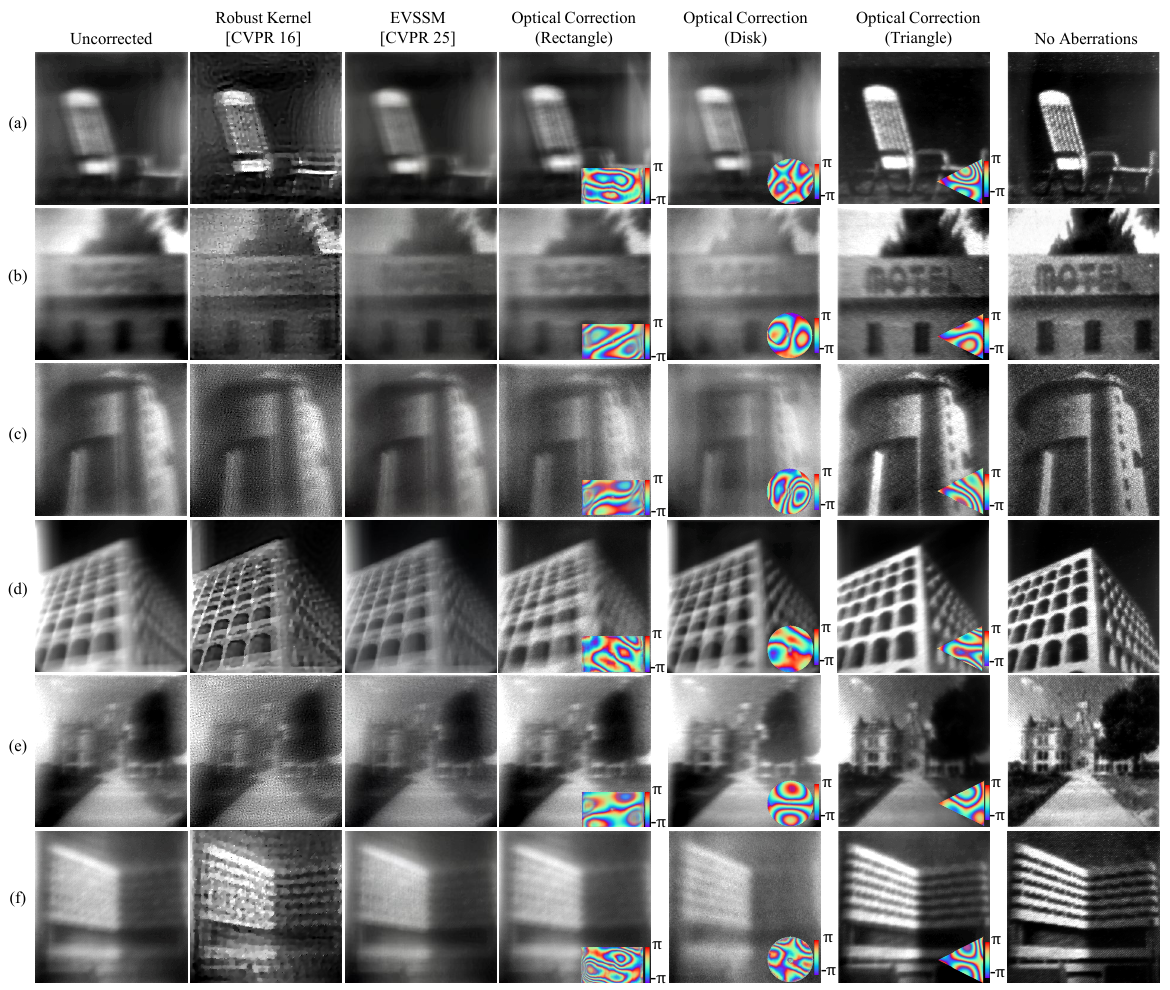}
    \caption{\textbf{Experimental results for guidestar-free imaging through real-world obscurants.}
    Experimental results demonstrate imaging through different obscurants: (a-b) nail polish, (c-d) onion skin, and (e-f) a $1^\circ$ diffuser. For each obscurant, columns show the uncorrected image, reconstructions by Robust Kernel~\cite{pan2016robust} and EVSSM~\cite{kong2025efficient}, optical corrections using rectangular, circular, and triangular apertures (with the corresponding estimated phase shown in the lower-right corner), and the no-aberration reference. Rectangular and circular apertures fail to recover fine details, while the triangular aperture leads to substantial improvement in image sharpness. Neither the state-of-the-art deep learning method EVSSM nor the traditional approach Robust Kernel is able to restore the image, highlighting the benefit of our method.}
    \Description{The figure is a grid of grayscale images with six rows 
    labeled (a) through (f) and seven columns. Each pair of rows corresponds 
    to a different obscurant: rows (a) and (b) use nail polish, rows (c) and 
    (d) use onion skin, and rows (e) and (f) use a one-degree diffuser. The 
    seven columns from left to right are: Uncorrected, Robust Kernel [CVPR 16], EVSSM [CVPR 25], Optical Correction (Rectangle), Optical 
    Correction (Disk), Optical Correction (Triangle), and No Aberrations. 
    The Uncorrected column shows heavily blurred images. EVSSM produces 
    results that remain blurred with no meaningful improvement. Robust Kernel 
    introduces severe blocky artifacts and distortions. The rectangular and 
    circular aperture optical corrections show some structure but fail to 
    recover fine details, with images remaining noticeably blurred. The 
    triangular aperture optical correction produces substantially sharper 
    images with clearly resolved scene content. Each optical correction image 
    includes a small inset in the lower-right corner showing the estimated 
    phase map, color-coded from negative pi to positive pi. The No 
    Aberrations column on the far right serves as the sharp reference.}
    \label{fig:guide_star_free_real}
\end{figure*}
\subsubsection{Ablation on PSF Estimation}
To investigate the impact of different PSF estimation methods on wavefront correction, we conduct an ablation study comparing several approaches. Fig.~\ref{fig:psf_ablation} shows qualitative results for three representative test scenes. For each scene, we present reconstructions obtained by applying wavefront correction using PSFs estimated by UFPNet~\cite{fang2023self}, Robust Kernel~\cite{pan2016robust}, our proposed method, and the ground-truth PSF. We also include the original uncorrected image and the reference image without aberrations.

The results indicate that using PSFs estimated by either UFPNet or Robust Kernel does not yield satisfactory wavefront correction. The restored images remain blurry and exhibit little improvement over the uncorrected input. In contrast, our estimated PSFs enable effective correction and produce reconstructions that are visually comparable to those achieved with the ground-truth PSF. This demonstrates that accurate PSF estimation is critical for successful wavefront correction and highlights the advantage of our proposed method over prior approaches that primarily focus on image deblurring rather than precise PSF recovery.

Quantitative results are summarized in Table~\ref{tab:ablation_psf}, which reports PSNR and SSIM for each method. Our PSF estimation pipeline achieves performance nearly matching that obtained with the ground-truth PSF.

\subsection{Experimental Results}
\noindent \textbf{Hardware Implementation.}
\label{sec:optical}
Figs.~\ref{fig:optical_setup} and \ref{fig:photo_optics} illustrate our experimental setup. The light source is a collimated 520~nm laser beam that passes through a polarizer and a rotating diffuser to provide spatial incoherence and reduce speckle. This beam illuminates the target, a natural scene. For natural scenes, we use physically printed images to ensure reproducible testing conditions. Lens L1 ($f=400$~mm) performs a Fourier transform of the target, and the obscurants (nail polish, onion skin, or a $1^\circ$ diffuser) are placed in the Fourier plane. Lenses L2 and L3 ($f=100$~mm each) form a $4f$ system that relays the aberration plane onto the HOLOEYE LETO-3 phase-only SLM, which corrects the distorted wavefront. Lens L4 ($f=100$~mm) then images the SLM output onto the Tucsen 20\,MP FL-20 sensor, resulting in an overall $4\times$ magnification 
reduction. To realize different apertures, we display a checkerboard pattern on the SLM at regions where zero amplitude is desired~\cite{mendoza2014encoding}. This setup adapts flexibly to different targets and magnifications.

To evaluate the generalization ability of our method, we conduct guidestar-free wavefront shaping experiments by imaging through real-world obscurants, including nail polish, onion skin, and $1^\circ$ diffusers. We compare three aperture types: triangular (asymmetric), circular, and rectangular (both symmetric). To ensure a fair comparison, we normalized optical flux by maintaining a constant aperture area of approximately 24~mm$^2$ for all shapes. The specific dimensions, constrained by the SLM's active area ($12.29 \times 6.91$ mm), are: rectangle $6.53 \times 3.67$ mm; triangle with base and height $6.91$ mm; and disk with radius $2.76$ mm. Each comparison uses the same physical obscurant placed at the same location in the optical path, with only the aperture shape varied, ensuring that observed performance differences are attributable solely to aperture geometry.
Testing images are randomly selected from a held-out subset of the MIT Places2 dataset~\cite{zhou2017places}, printed on transparent film, and positioned at the target plane as shown in Figs.~\ref{fig:optical_setup} and \ref{fig:photo_optics}. 

As illustrated in Fig.~\ref{fig:guide_star_free_real}, wavefront correction with the asymmetric triangular aperture is significantly more effective than with symmetric rectangular or circular apertures. Furthermore, the figure demonstrates that our method achieves robust optical correction even where state-of-the-art image deblurring algorithms fail. These quantitative comparisons are summarized in Table~\ref{tab:quan_results}. For a comprehensive evaluation, we detail the full PSNR and SSIM metrics for each aperture type in Table~\ref{tab:real_psnr} (Appendix~\ref{appendix:quan_real}). We evaluated ten scenes per obscurant type, utilizing distinct unknown obscurants in every experiment. These extended results consistently confirm that the asymmetric triangular aperture yields the best overall correction performance.
\section{Discussion}
\label{sec:Discussion}

In summary, we have introduced a powerful real-time closed-loop AO framework that operates without guidestars, coded measurements, or specialized wavefront sensors. Our system combines deep learning and asymmetric apertures to enable high-quality imaging through severe unknown real-world obscurants. Compared to existing guidestar-free wavefront shaping and image deblurring methods, our approach consistently delivers superior image restoration quality while requiring significantly fewer measurements and far less computation. By removing the need for guidestars, this work significantly broadens the applicability of AO to settings where guidestars are impractical or unavailable.

\paragraph{Limitations and Future Work.}  
While our method significantly improves imaging quality, residual higher order aberrations---that our system was not trained to estimate or correct---still persist. Additionally, our present system is restricted to correcting spatially invariant isoplanatic optical aberrations, i.e.,~aberrations within the memory effect region. 
One could digitally correct anisoplanatic aberrations with our system by imaging patch-by-patch through a scene and stitching together the results. 
Alternatively, one could optically correct anisoplanatic aberrations with a multi-conjugate extension of our approach---such an extension represents an exciting direction for future work. 
Likewise, while our guidestar-free aberration correction technique is computationally lightweight, our current system was not optimized for latency. At present, I/O bottlenecks between the SLM, camera, and computer limit our system's refresh rate to 1 frame per second (FPS). These bottlenecks are not fundamental in nature and, with additional engineering effort, our system could run far faster.

\begin{acks}
This work was supported in part by NIH award no.~R01DE032051 and R21EY037446A, ONR award nos.~N000142312752 and N000142312714, ARO ECP award no.~W911NF2420113, and NSF CAREER award no.~2339616.
\end{acks}

\bibliographystyle{ACM-Reference-Format}
\bibliography{sample-base}

\appendix
\begin{figure*}[!h]
   \includegraphics[width=1.0\linewidth]{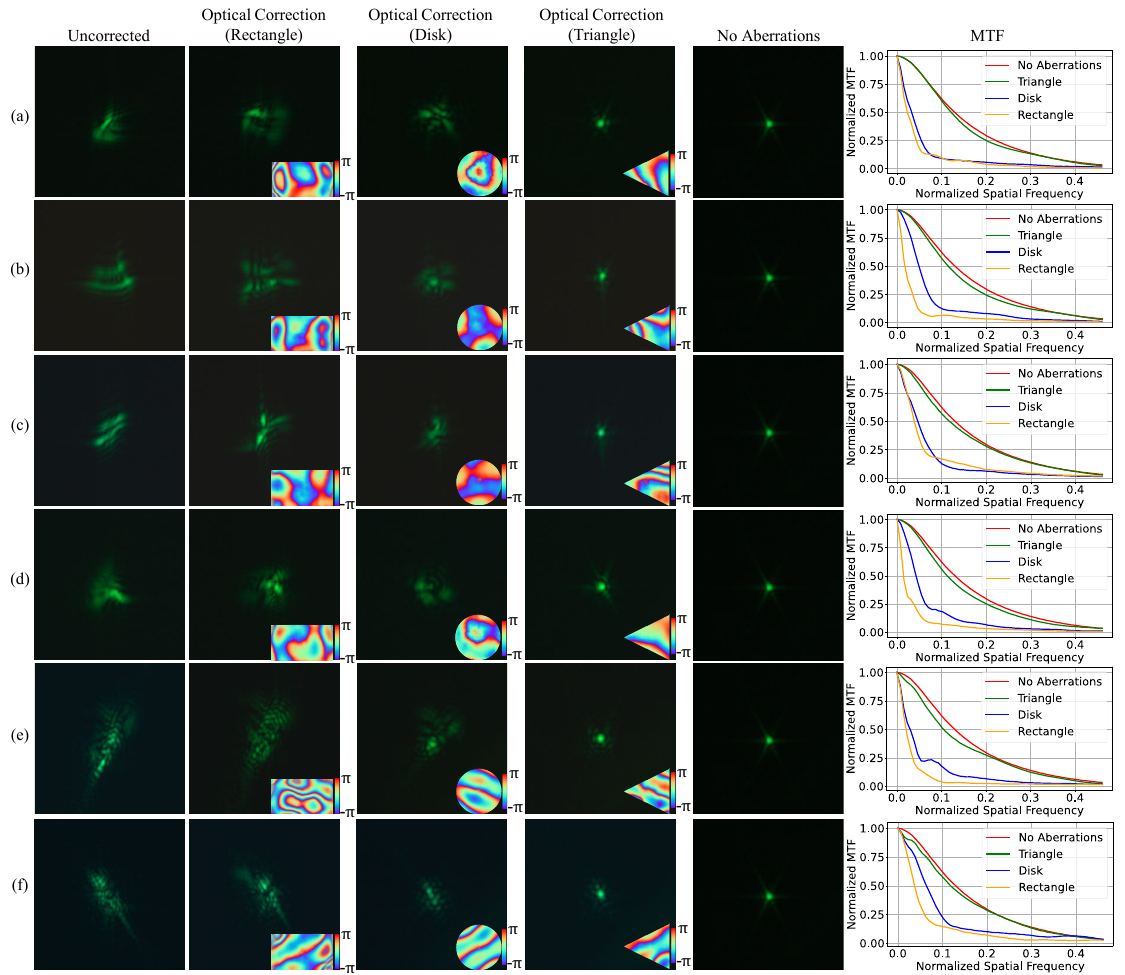}
   \caption{\textbf{Guidestar-based experimental validation of Phase U-Net generalization.} 
   Experimental pinhole imaging through (a–b) nail polish, (c–d) onion skin, and (e–f) a $1^\circ$ diffuser. Columns show the uncorrected image, corrected results using rectangular, circular, and triangular apertures (with estimated phase at lower right), the no-aberration reference captured with the triangular aperture under diffraction-limited conditions, and MTF plots. Although trained only on Zernike-based aberrations, the Phase U-Net corrects real aberrations effectively.}
   \Description{The figure is a grid with six rows labeled (a) through (f) 
    and six columns. Each pair of rows corresponds to a different obscurant: 
    rows (a) and (b) use nail polish, rows (c) and (d) use onion skin, and 
    rows (e) and (f) use a one-degree diffuser. The first five columns show 
    green-tinted pinhole images on dark backgrounds. The columns from left to 
    right are: Uncorrected, Optical Correction (Rectangle), Optical 
    Correction (Circle), Optical Correction (Triangle), and No Aberrations. 
    The Uncorrected column shows spread-out, distorted pinhole images. The 
    rectangular and circular aperture corrections show partially concentrated 
    light but with residual spread. The triangular aperture correction 
    produces the most compact pinhole images, closest to the No Aberrations 
    reference which shows a tight, concentrated spot. Each optical correction 
    image includes a small inset in the lower-right corner showing the 
    estimated phase map, color-coded from negative pi to positive pi. The 
    sixth column contains MTF plots for each row, with normalized MTF on the 
    vertical axis and normalized spatial frequency on the horizontal axis. 
    Each plot shows four curves: No Aberrations (red), Triangle (green), 
    Disk (blue), and Rectangle (orange). Across all rows, the Triangle curve 
    closely follows the No Aberrations curve, while the Disk and Rectangle 
    curves fall off more rapidly at higher spatial frequencies.}
   \label{fig:guide_star_real}
\end{figure*}
\begin{figure*}[t]
    \centering
    \includegraphics[width=0.8\linewidth]{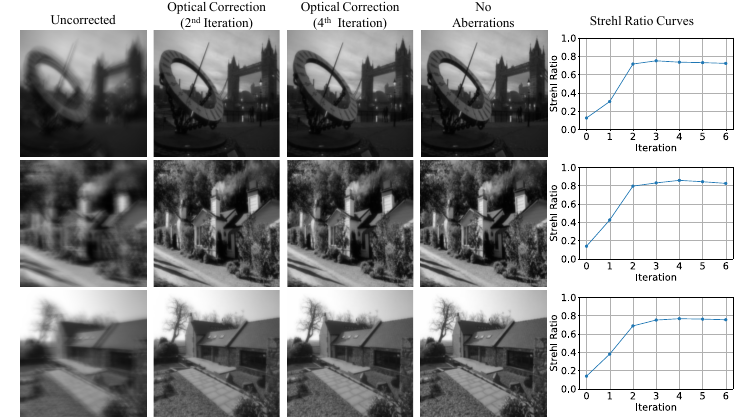}
    \caption{\textbf{Strehl ratio across iterations.} Most optical aberrations are corrected within the first two iterations, with subsequent iterations providing smaller improvements. The Strehl ratio---defined as the peak intensity of the corrected wavefront normalized by the peak intensity of an ideal diffraction-limited system---quantifies the effectiveness of aberration correction. Higher Strehl ratios correspond to improved image quality. This motivates us to use four iterations throughout our experiments.}
    \Description{A grid of three rows, each showing a different natural scene: 
    a sculpture with Tower Bridge in the background, a house with ivy and 
    trees, and a rural building with a tiled roof. Each row contains four 
    grayscale images and one line plot. The four images show, from left to 
    right: the uncorrected blurry image, optical correction after the 2nd 
    iteration, optical correction after the 4th iteration, and a 
    no-aberrations reference. The uncorrected images appear noticeably 
    blurred, the 2nd iteration images show substantial improvement, and the 
    4th iteration images are visually close to the no-aberrations reference. 
    The rightmost column shows a Strehl ratio curve plotted over iterations 0 
    to 6, with Strehl Ratio on the vertical axis ranging from 0 to 1 and 
    Iteration on the horizontal axis. In all three rows, the Strehl ratio 
    starts near 0.1, rises steeply by iteration 2 to around 0.4, reaches 
    approximately 0.7 to 0.85 by iteration 3, and then plateaus through 
    iteration 6, demonstrating that most aberration correction occurs within 
    the first few iterations.}
    \label{fig:strehl_ratio}
\end{figure*}
\section*{Appendices}
\section{Guidestar-Based Real-World Experiments}
While our primary contribution is a guidestar-free adaptive optics pipeline, we also provide supplementary experiments to demonstrate that our method is effective when a guidestar is available. In these experiments, instead of using our proposed PSF U-Net to estimate the PSF from a blurred measurement, we directly use the PSF captured with a pinhole guidestar and feed it to the Phase U-Net for phase recovery. This setup bypasses the PSF estimation step and is designed to directly validate the performance of our phase U-Net when the ground truth PSF is available. These experiments are analogous to those performed concurrently in~\cite{chimitt2025wavefront}. 

Beyond qualitative comparisons of PSF corrections in Fig.~\ref{fig:guide_star_real}, we quantitatively assess performance by analyzing the modulation transfer function (MTF) for all three aperture shapes. The MTF characterizes the system’s ability to preserve image contrast across spatial frequencies and is a standard metric for evaluating imaging resolution and quality~\cite{sun2021end, tian2015quantitative, haefner2018mtf}. The results confirm that our asymmetric triangular aperture achieves a higher MTF compared to conventional symmetric designs, further validating its effectiveness in wavefront correction.

It is important to note that both the PSF and phase used for training the Phase U-Net are generated from Zernike aberration profiles. Therefore, these real-world experiments further demonstrate that our Phase U-Net generalizes well beyond Zernike-based synthetic data and performs robustly under real, complex aberrations introduced by real-world obscurants.

\begin{table}[!t]
\caption{\textbf{Quantitative evaluation of PSNR / SSIM on guidestar-free real-world experiments under different obscurants and apertures.} 
We test 10 scenes for each obscurant type (nail polish, onion skin, and optical diffusers). Obscurants are distinct in every experiment. Three apertures are evaluated and the results indicate that the triangular aperture exhibits superior performance. \textbf{Bold} indicates the best result.}
\label{tab:real_psnr}
\begin{tabular}{@{}lccc@{}}
\toprule
\multirow{2}{*}{Aperture} & \multicolumn{3}{c}{Obscurant} \\ \cmidrule(l){2-4} 
                           & Nail Polish             & Onion Skin              & Optical Diffuser               \\ \midrule
Rectangle                  & 14.98 / 0.5290          & 15.65 / 0.4849          & 15.59 / 0.4852          \\
Disk                       & 15.36 / 0.5543          & 15.13 / 0.4572          & 14.99 / 0.4791          \\
Triangle                   & \textbf{25.51 / 0.7286} & \textbf{25.82 / 0.7885} & \textbf{24.67 / 0.7093} \\ \bottomrule
\end{tabular}
\end{table}

\section{Reproducibility}
To facilitate reproducibility, we have released our training code, data generation scripts, and pretrained models for direct use in baseline comparisons.

\section{Strehl Ratio–Based Iteration Selection}
\label{sec:strehl}

To evaluate the convergence of our AO pipeline, we monitor the Strehl ratio across iterations. As shown in Fig.~\ref{fig:strehl_ratio}, the Strehl ratio improves rapidly in the first two iterations, indicating that most aberrations are corrected early in the process. Additional iterations provide smaller improvements, gradually reducing residual errors. Based on this analysis, we fix the number of iterations to four throughout our experiments to ensure both performance and efficiency.

\section{Quantitative Evaluation of PSNR and SSIM on Guidestar-Free Real-World Experiments Under Different Obscurants and Apertures}
\label{appendix:quan_real}
Table~\ref{tab:real_psnr} demonstrates that the asymmetric triangular aperture outperforms the other two symmetric apertures across all three obscurants, highlighting its superior capability for guidestar-free imaging through real-world obscurants.
\begin{figure*}[!t]
    \centering
    \includegraphics[width=\textwidth]{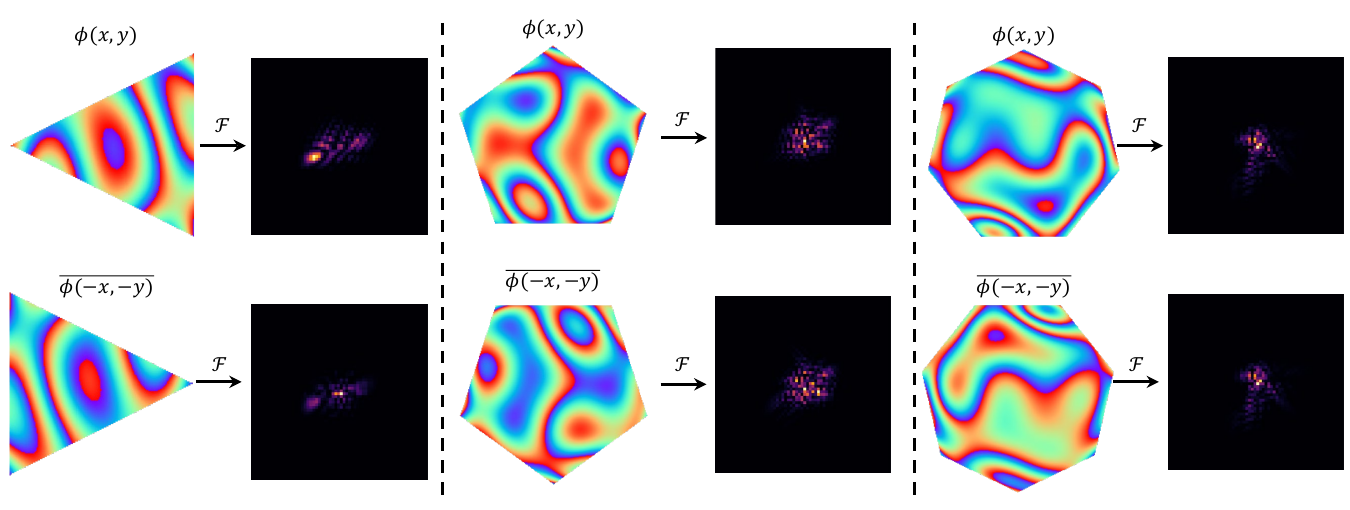}
    \caption{\textbf{Comparison of PSFs generated by triangular, pentagonal, and heptagonal apertures and their conjugate flips.} The triangular aperture produces clearly distinguishable PSF pairs, while pentagonal and heptagonal apertures yield very similar conjugate PSFs.}
    \Description{A two-row, six-column grid divided into three sections by vertical dashed lines, one section per aperture shape: triangular (left), pentagonal (center), and heptagonal (right). Each section shows two rows: the top row displays the original phase map phi(x,y) on the aperture followed by a Fourier transform arrow and the resulting PSF; the bottom row shows the conjugate-flipped phase map phi-bar(-x,-y) on the same aperture shape and its resulting PSF. The phase maps are color-coded from blue to red. For the triangular aperture, the two PSFs are visibly distinct, enabling unambiguous phase retrieval. For the pentagonal and heptagonal apertures, the two PSFs in each section appear very similar, demonstrating that these apertures do not sufficiently break the conjugate-flip symmetry and are therefore less suitable for phase retrieval.}
    \label{fig:toy_penta}
\end{figure*}
\begin{figure}[!t]
    \centering
    \includegraphics[width=1.0\linewidth]{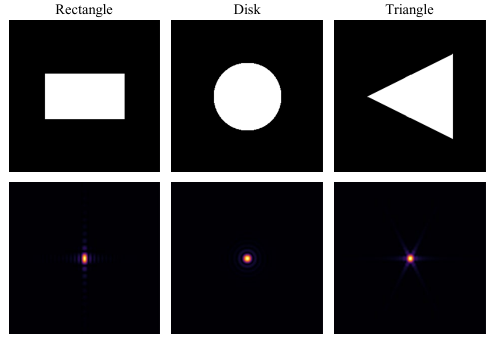}
    \caption{\textbf{Shape-dependent diffraction-limited PSFs for equal-area apertures.} Top row: rectangular, circular, and triangular apertures, all with equal area ($\approx 24\ \text{mm}^2$). Bottom row: corresponding simulated diffraction-limited PSFs. The rectangular aperture produces a sinc-like cross pattern, the circular aperture produces isotropic Airy-like rings, and the triangular aperture produces a 6-spike diffraction pattern. Despite these differences in side-lobe structure, the central lobes remain largely consistent across all three apertures, ensuring that performance comparisons are not confounded by differences in diffraction-limited image quality.}
    \Description{Two rows of images showing three aperture shapes and their corresponding diffraction-limited PSFs. The top row displays rectangular, circular, and triangular apertures, all with equal area. The bottom row shows the simulated PSFs: the rectangular aperture produces a sinc-like cross pattern, the circular aperture produces isotropic Airy-like rings, and the triangular aperture produces a six-spike diffraction pattern.}
    \label{fig:diffraction_psfs}
\end{figure}

\section{Choice of Aperture Geometry}
The use of a triangular aperture in our method is motivated by the uniqueness conditions established by \citet{chimitt2025wavefront}, who showed that asymmetric pupils can resolve the conjugate flip ambiguity in wavefront estimation. Specifically, symmetric apertures such as circles and squares produce two equally good global minima in the optimization landscape, making it impossible to distinguish the correct solution from its conjugate flip. Asymmetric apertures such as triangles and pentagons break this symmetry. Among these, the triangle produces a single global minimum corresponding to the correct solution, whereas the pentagon produces two local minima with different objective values that a well-designed algorithm can distinguish~\cite{chimitt2025wavefront}.
\begin{figure}[!t]
\includegraphics[width=1.0\linewidth]{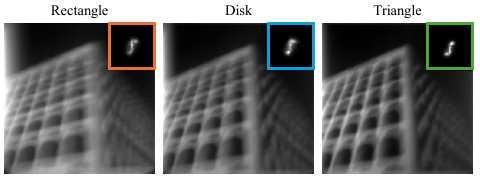}
    \caption{\textbf{PSF U-Nets perform equally well across different apertures.} The performance of PSF U-Net remains consistent for rectangular, circular, and triangular apertures under the same obscurant conditions. This indicates that the improved optical correction observed with asymmetric triangular apertures depends solely on whether the Phase U-Net can resolve ambiguities and retrieve the unique phase.}
    \Description{
    The figure shows three grayscale images of a building, each corresponding to a different aperture shape: rectangle, disk, and triangle. In each panel, a smaller inset image is located at the upper right corner and outlined in a colored border. The insets display the estimated point spread function (PSF) for each aperture type. The main images appear similarly blurred, indicating that the PSF U-Net performs consistently across different aperture shapes. The insets further highlight that the quality of the estimated PSF is similar for all three aperture geometries.
    }
    \label{fig:psf_performance}
\end{figure}
To verify this in the context of our method, we tested pentagonal and heptagonal apertures in addition to the triangular aperture. As shown in Fig.~\ref{fig:toy_penta}, the PSFs produced by a pentagonal aperture and its conjugate flip are very similar, with only subtle differences that lead to potential ambiguity in phase retrieval. The same issue arises with heptagonal apertures. The triangular aperture, by contrast, produces sufficiently distinct PSFs between conjugate pairs, making it the most reliable choice among the apertures we tested.

\section{Diffraction-Limited PSF Comparison of Rectangular, Circular, and Triangular Apertures with Equal Area}
Fig.~\ref{fig:diffraction_psfs} shows the simulated diffraction-limited PSFs for the three aperture shapes considered in this paper---rectangular, circular, and triangular ---all with equal area ($\approx 24\ \text{mm}^2$). Each aperture produces a characteristic diffraction pattern: the rectangular aperture produces a sinc-like cross pattern, the circular aperture yields isotropic Airy-like rings, and the triangular aperture yields a six-spike diffraction pattern. Importantly, despite these shape-dependent differences in the side-lobe structure, the central lobe of the PSF---which carries the majority of the energy---remains largely consistent across the three apertures. This ensures that the performance comparisons presented in this paper are not confounded by differences in diffraction-limited image quality, and that any observed advantages of the triangular aperture can be attributed to its lack of point symmetry rather than to differences in the underlying diffraction-limited PSF.
\begin{figure*}[!t]
    \centering
    \includegraphics[width=0.8\linewidth]{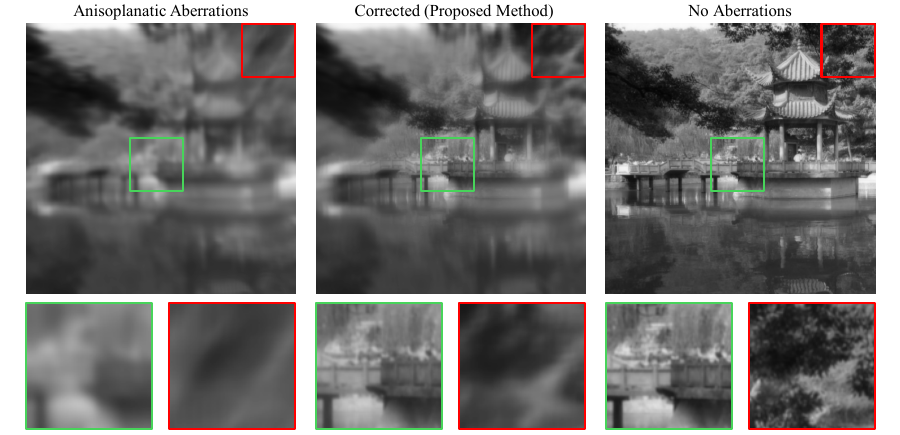}
    \caption{\textbf{Reconstruction quality within and beyond the memory effect region.} From left to right: image degraded by anisoplanatic aberrations, result after correction using our proposed method, and aberration-free reference. The PSF is estimated from the central isoplanatic region. Zoomed-in patches highlight two representative regions: the green box marks a central region within the memory effect range, where the correction is accurate; the red box marks a peripheral region beyond the memory effect range, where residual blur and artifacts remain. The method degrades gradually, producing localized artifacts at the periphery rather than a global correction failure.}
    \Description{Three images shown side by side: the leftmost shows a scene degraded by anisoplanatic aberrations, the middle shows the result after correction using the proposed method, and the rightmost shows the aberration-free reference. Two zoomed-in patches are highlighted: a green box marking a central region within the memory effect range where correction is accurate, and a red box marking a peripheral region beyond the memory effect range where residual blur and artifacts remain.}
    \label{fig:beyond_memory_effect}
\end{figure*}
\section{Aperture Shape Does Not Affect PSF Estimation}
To assess the influence of aperture shape on PSF U-Net performance, we evaluate its accuracy with triangular, rectangular, and circular apertures using the same physical obscurant placed at the same location in the optical path, with only the aperture pattern on the SLM varied (Fig.~\ref{fig:psf_performance}). The consistent results across different aperture shapes demonstrate that PSF U-Net provides reliable PSF estimation regardless of aperture geometry, and does not limit the overall effectiveness of optical correction.

\section{Corrections Beyond the Memory Effect Region}
In the analysis presented thus far, we have assumed that the scene lies within the memory effect region, where a single phase screen adequately describes the aberrations. In practice, however, scenes may extend beyond this range, resulting in anisoplanatic aberrations that vary spatially across the field of view. It is therefore important to understand how the proposed method performs when this assumption is violated.

To investigate this, we simulated a scenario with spatially varying aberrations using Fresnel propagation across multiple phase screens, each representing an aberration layer at a distinct depth within the obscurant. Fig.~\ref{fig:beyond_memory_effect} shows the degraded 
image with anisoplanatic aberrations, the result after correction using our proposed 
method, and the aberration-free reference. The PSF was estimated from the central 
region of the image (indicated by the green box), which lies 
within the memory effect range. The correction remains effective within this region, 
while artifacts appear in peripheral areas (indicated by the red box) 
where the local aberration profile deviates from the assumed isoplanatic model. 
Notably, the reconstruction does not fail globally; rather, it degrades gradually 
beyond the memory effect region. This suggests that anisoplanatic aberrations could in principle be corrected by applying our method patch-by-patch across the scene and stitching together the results.

\begin{table}[!t]
\centering
\caption{\textbf{Model performance across an increasing number of Zernike modes.} The model is trained exclusively on wavefront aberrations parameterized by 36 Zernike modes (in-distribution). At test time, we evaluate generalization by increasing the number of modes to 55 and 78 (out-of-distribution), introducing higher-order aberration components unseen during training. Performance degrades progressively as the complexity of the aberration exceeds the training distribution.}
\label{tab:zernike_scaling}
\begin{tabular}{@{}lcc@{}}
\toprule
Test Setting & PSNR $\uparrow$ & SSIM $\uparrow$ \\ \midrule
In-Distribution (36 modes)      & 37.95 & 0.9865 \\
Out-of-Distribution (55 modes)  & 31.06 & 0.9530 \\
Out-of-Distribution (78 modes)  & 22.50 & 0.7288 \\ \bottomrule
\end{tabular}
\end{table}
\section{Robustness to Higher-Order Zernike Modes}

To evaluate the robustness of the proposed method to higher-order aberrations beyond the training distribution, we tested the model on aberrations composed of 55 and 78 Zernike modes, while the model was trained using 36 modes. Table~\ref{tab:zernike_scaling} summarizes the results. The model degrades gracefully at 55 modes but shows a more significant performance drop at 78 modes, suggesting that retraining or fine-tuning on higher-order modes would be necessary for applications where such aberrations are dominant.

It should be noted that truly discontinuous or pixel-wise random phase distortions, such as those arising from strong atmospheric turbulence or rough optical surfaces, may not be well represented by any finite set of Zernike modes. Addressing such scenarios would likely require an alternative basis, such as Kolmogorov-based phase screens for atmospheric turbulence, or training directly on representative phase realizations. We consider this an important direction for future work.

\section{Generalization to New Optical Configurations}
\label{supp:generalization}
The robustness of our deep neural network-based pipeline relies on the high fidelity of the forward physical model used for training data synthesis. In principle, our physics-informed simulation strategy avoids overfitting to a specific hardware instance and enables generalization across diverse optical setups by aligning simulated priors with target system constraints.
To adapt the pipeline to a new hardware configuration, we calibrate the forward model using two primary physical parameters. First, magnification and pixel pitch define the spatial sampling of the PSF on the sensor; by scaling the magnification factor, the simulator ensures the PSF's spatial support matches the effective pixel size of the new hardware. Second, the numerical aperture (NA) dictates the diffraction-limited frequency support of the system. We account for this by updating the bandwidth of the pupil function in our simulations, ensuring that the generated PSFs accurately represent the angular resolution and frequency cutoff of the specific optical setup.

To validate this transition, a one-time calibration measurement of a point source (e.g., a pinhole) is performed. The simulated PSF is then matched to the experimental PSF's full width at half maximum (FWHM) and side-lobe intensity to ensure high-fidelity corrections.

\end{document}